\documentclass[twocolumn,pra,superscriptaddress,amssymb,amsmath,amsmath,floatfix]{revtex4-1}
\usepackage{graphicx}
\usepackage{epstopdf}
\usepackage{comment}
\usepackage{bm}
\usepackage{wrapfig}
\usepackage{times}
\usepackage[colorlinks=true,linkcolor=blue,urlcolor=blue,citecolor=blue]{hyperref}
\usepackage{float}

\begin{document}
\title{Diatomic molecules of alkali-metal and alkaline-earth-metal atoms: interaction potentials, \\ dipole moments, and polarizabilities}

\author{Hela Ladjimi}
\affiliation{Faculty of Physics, University of Warsaw, Pasteura 5, 02-093 Warsaw, Poland}
\author{Micha{\l} Tomza}
\email{michal.tomza@fuw.edu.pl}
\affiliation{Faculty of Physics, University of Warsaw, Pasteura 5, 02-093 Warsaw, Poland}

\date{\today}

\begin{abstract}
Ultracold diatomic molecules find application in quantum studies ranging from controlled chemistry and precision measurement physics to quantum many-body simulation and potentially quantum computing. Accurate knowledge of molecular properties is required to guide and explain ongoing experiments. Here, in an extensive and comparative study, we theoretically investigate the electronic properties of the ground-state diatomic molecules composed of alkali-metal (Li, Na, K, Rb, Cs, Fr) and alkaline-earth-metal (Be, Mg, Ca, Sr, Ba, Ra) atoms. We study 78 hetero- and homonuclear diatomic combinations, including 21 alkali-metal molecules in the $X^1\Sigma^+$ and $a^3\Sigma^+$ electronic states, 36 alkali-metal--alkaline-earth-metal molecules in the $X^2\Sigma^+$ electronic state, and 21 alkaline-earth-metal molecules in the $X^1\Sigma^+$ electronic state. We calculate potential energy curves, permanent electric dipole moments, and polarizabilities using the hierarchy of coupled cluster methods upto CCSDTQ with large Gaussian basis sets and small-core relativistic energy-consistent pseudopotentials. We collect and analyze corresponding spectroscopic constants. We estimate computational uncertainties and compare the present values with previous experimental and theoretical data to establish a new theoretical benchmark. The presented results should be useful for further application of the studied molecules in modern ultracold physics and chemistry experiments.
\end{abstract}

\maketitle

\section{Introduction}

Cold and ultracold molecules offer exciting prospects for fundamental quantum physics and physical chemistry studies as well as for new quantum technology developments~\cite{CarrNJP09}. Within the past two decades, ultracold polar molecules have been established as great candidates for a plethora of applications ranging from ultracold controlled chemistry~\cite{QuemenerCR12,BohnScience17} and precision measurements of fundamental constants~\cite{SafronovaRMP18,DeMilleScience17} to quantum simulation of many-body systems~\cite{MicheliNatPhys06,LahayeCPP09} and quantum computation~\cite{DeMillePRL02,NiCS18}. One of the unique features of ultracold polar molecules is the possibility of controlling their states and intermolecular dipolar interactions with an external electric field~\cite{GadwayJPB16}. The achieved exquisite control of both molecular internal quantum states and external motion, enabled by ultralow temperatures~\cite{QuemenerCR12}, propels further intensive theoretical and experimental investigations.

Ultracold molecules can be produced either directly using laser~\cite{ShumanNature10,BarryNature14,TruppeNP17}, evaporative~\cite{StuhlNature12,ValtolinaNature20,SchindewolfNature22,Bigagli2023}, or sympathetic~\cite{SonNature19,ParkNP23} cooling or indirectly by associating pre-cooled atoms~\cite{KohlerRMP06,JonesRMP06}. However, ultracold diatomic molecules of alkali-metal and alkaline-earth-metal atoms have only been produced using indirect methods. Fortunately, alkali-metal and alkaline-earth-metal atoms can be favorably laser cooled to ultralow temperatures. Ultracold gases of alkali-metal molecules such as KRb~\cite{NiScience08,HuScience19}, RbCs~\cite{TakekoshiPRL14,MolonyPRL14}, NaK~\cite{ParkPRL15,LiuPRA19,VogesPRL20,BausePRA21}, NaRb~\cite{GuoPRL16}, NaCs~\cite{StevensonPRL23}, and LiK~\cite{He2023} in the ground rovibrational level of the singlet electronic state were produced from ultracold atoms by magnetoassociation using magnetic Feshbach resonances~\cite{KohlerRMP06,ChinRMP10} followed by an optical stabilization using the Stimulate Adiabatic Raman Passage (STIRAP)~\cite{JonesRMP06,VitanovRMP17}. A similar scheme was used to obtain ultracold gases of Rb$_2$~\cite{DenschlagPRL08}, Cs$_2$~\cite{LangPRL08,DanzlScience08}, NaLi~\cite{RvachovPRL17}, and Li$_2$~\cite{PolovyPRA20} molecules in the ground rovibrational level of the lowest triplet electronic state. Alternatively, molecules such as RbCs~\cite{SagePRL05}, Cs$_2$~\cite{ViteauScience08}, LiCs~\cite{DeiglmayrPRL08}, and recently Sr$_2$~\cite{StellmerPRL12,LeungNJP21} were produced in the rovibrational ground state using all-optical photoassociation schemes. Degenerate Fermi gases of polar KRb~\cite{DeMarcoScience19} and NaK~\cite{DudaNP23} molecules were formed. The Bose-Einstein condensation of weakly bound Li$_2$, K$_2$, and Cs$_2$ Feshbach molecules was realized~\cite{JochimScience03,GreinerNature03,ZhangNature21}, and very recently, the condensation of deeply-bound ground-state NaCs molecules was obtained~\cite{Bigagli2024}. First steps toward producing ultracold SrRb~\cite{BarbeNP18} and SrLi~\cite{ZhuPRA20} molecules using narrow magnetic Feshbach resonances were achieved. Finally, the formation of ultracold NaCs~\cite{LiuScience18,ZhangPRL20,YuPRX21,Cairncross2021}, Rb$_2$~\cite{XiaodongScience20}, and RbCs~\cite{Ruttley2023} molecules was possible at the single-molecule level using optical tweezers~\cite{KaufmanNP21}. 

Ultracold quantum-controlled chemical reactions were studied in pioneering experiments with KRb molecules~\cite{OspelkausScience10,NiNature10,MirandaNatPhys11,HuScience19,LiuNP20,LiuNature20,HuNC21,WilliamScience22,Liu2023}, followed by investigations involving Rb$_2$~\cite{DrewsNC17}, NaRb~\cite{GuoPRX18,YeSA18,GersemaPRL21}, NaCs~\cite{GregoryNC19,GregoryPRL20}, and NaK~\cite{BausePRR21,GersemaPRL21} species. Suppressing chemical reactivity and short-range losses~\cite{MaylePRA12,MaylePRA13,ChristianenPRL19,JachymskiPRA22,BauseJPCA23} was realized by shielding with an electric field for KRb~\cite{MatsudaScience20} and with a microwave field for and NaK~\cite{SchindewolfNature22}, NaCs~\cite{Bigagli2023}, and NaRb~\cite{Lin2023} molecules. Feshbach resonances controlled with a magnetic field were demonstrated in ultracold NaK+K~\cite{YangScience19,SuPRL22}, NaLi+Na~\cite{SonScience22,Park2023,KarmanPRA23}, and NaLi+NaLi~\cite{ParkNature22} collisions and used to create ultracold weakly-bound NaK$_2$ triatomic molecules from a NaK+K mixture~\cite{YangNature22,YangScience22,CaoPRL24}. Ultracold KRb molecules were trapped in a three-dimensional optical lattice, and dipolar spin-exchange interactions between lattice-confined molecules were observed~\cite{BoNature13} and tuned~\cite{LiNature23}. The first realization of a molecular quantum gas microscope with NaRb molecules~\cite{RosenbergNP22,Christakis2022} was also reported. Long-lived coherence of molecular qubits based on ultracold NaK~\cite{ParkScience17} and RbCs~\cite{GregoryNP21,GregoryNP24} molecules was shown. Ultracold KRb molecules were employed for precision measurements of the variation of the electron-to-proton mass ratio~\cite{KobayashiNC19}, while ultracold Sr$_2$ molecules in an optical lattice were established as a molecular clock for metrology and probing the fundamental laws of nature~\cite{McGuyerNP15,McDonaldNature16,KondovNP19,LeungPRX23}.

Molecular formation and application described above would not be possible without preceding detailed experimental spectroscopic studies and extensive theoretical \textit{ab initio} electronic structure calculations of underlying molecular properties. Different applications require data at different levels of accuracy. Accurate measurements can ultimately, in most cases, provide more accurate results than theoretical computations. Nevertheless, \textit{ab initio} quantum-chemical calculations of potential energy curves, permanent and transition electric dipole moments, and fine and hyperfine couplings, used next in rovibrational and scattering calculations, are often essential to guide and explain experimental efforts.

\begin{table}[tb!]
\caption{References to previous experimental and theoretical works on the potential energy curves of the alkali-metal diatomic molecules in the $X^1\Sigma^+$ or $a^3\Sigma^+$ electronic state.} 
\label{tab:Ref_AM-AM}
\begin{ruledtabular}
\begin{tabular}{lll}
Molecule & Experiment & Theory \\
\hline
Li$_2$ 
  & \cite{VelascoJCP69,HesselJCP79,VermaJMS81,VermaJCP83,VergesCPL83,XieJCP85,BarakatCP86,MartinSAA88,LintonJCP89,AbrahamJCP95,LintonJMS99,ZavitsasJMS03,CoxonJMS06,LeRoyJCP09,DattaniJMS11,SemczukPRA13,SemczukPRL14} 
  &  \cite{RoachJMS72,WatsonCPL77,OlsonCP77,KonowalowJCP79,MaynauCPL81,JonssonJCP81,DaviesCPL81,MullerJCP84,KonowalowCP84,SchmidtCP85,IgelJCP86,KaldorCP90,SunCPL92,BoldyrevJCP93,PoteauJMS95,JasikCP06,NoroTCA08,MusialJCP13,MusialJCTC14,NasiriCPL15,BaryszJCTC16,LesiukPRA20,QiCTP21,NakatsujiJCP22,ShengPCCP22}\\
LiNa 
  & \cite{EngelkeCP82,FellowsMP88,SteinkePRA12,FellowsJCP91,SteinkePRA12,RvachovPCCP18}
  &  \cite{RoachJMS72,MaynauCPL81,DaviesCPL81,MullerJCP84,IgelJCP86,SchmidtCPL84,BoldyrevJCP93}\\
  & & \cite{AymarJCP05,NeogradyCCCC05,PetsalakisJCP08,KhelifiJRLR09,MabroukJPB09,FedorovJCP14,BellayouniAQC14,MieszczaninMP14,MabroukMP20,GronowskiPRA20,MusialJCP21} \\
LiK 
  & \cite{EngelkeCP84,BednarskaJCP97,BednarskaJMS98,MartinJCP01,SalamiJCP07,TiemannPRA09} 
  & \cite{RoachJMS72,MaynauCPL81,MullerJCP84,IgelJCP86,RousseauCP99,AymarJCP05,MiadowiczOP13}\\
  &  & \cite{FedorovJCP14,BellayouniAQC14,XiaoJQSRT13,MusialAQC16b} \\
LiRb  
  & \cite{IvanovaJCP11a,DuttaCPL11,StevensonPRA16} 
  & \cite{IgelJCP86,KorekCP00,AymarJCP05,JendoubiJPCA12,DardouriIJQC12,FedorovJCP14,BellayouniAQC14,YouPCCP16,YouSAA16,SkrzynskiMolecules23} \\
LiCs 
  & \cite{StaanumPRA07} 
  & \cite{IgelJCP86,KorekCJP00,AymarJCP05,DardouriIJQC12,BellayouniAQC14,FedorovJCP14,MabroukJPCA10} \\
LiFr 
  & -
  & \cite{BellayouniAQC14,JendoubiAJCE22,ShundalauJQSRT23} \\
Na$_2$
  & \cite{DemtroderJCP69,KuschJCP78,MaynauCPL81,KatoJCP82,BarrowCPL84,LuhJMS85,LiJCP85,JonesPRA96,SamuelisPRA00,HoJCP00,IvanovJCP03,CamachoJPB05,LauJCP16,KnoopPRA11,BauerJCP19} 
  &  \cite{RoachJMS72,StevensJCP77,KonowalowJCP80,ValancePLA81,DaviesCPL81,ValanceJPB82,PartridgeJCP83,JeungJPB83b,MullerJCP84,MagnierJCP93,IgelJCP86,NeogradyCCCC05}\\ & & \cite{NoroTCA08,MusialJCP13,BaryszJCTC16,ZhangCJP16,AdamsonRJPCB20,MusialMP22}\\
NaK  
  & \cite{BrefordJCP79,EiselJCP79,WormsbecherJCP81,RossMP85,KowalczykJCP89,IshikawaJCP94,KrouJMS98,FerberJCP00,RussierJPB00,GerdesEPJD08,GerdesEPJD11,TemelkovPRA15,HartmannPRA19}
  & \cite{RoachJMS72,JanoschekCPL78,MaynauCPL81,JeungCPL83,MullerJCP84,StevensJCP84,IgelJCP86,MagnierPRA96,MagnierJMS00,AymarJCP05}\\
  & & \cite{NeogradyCCCC05,AymarMP07,AlloucheJCP11,FedorovJCP14,MusialAQC16,ZhangCJCP19,SegoviaMP19,NakanoJCTC19,BaczekMP22} \\
NaRb 
  & \cite{TakahashiJCP81,WangJCP91,KasaharaJCP96,ZemkeJCP01,DocenkoPRA02,DocenkoPRA04,PashovPRA05,ParkNJP15}
  & \cite{IgelJCP86,KorekCP00,ZaitsevskiiPRA01,AymarJCP05,NeogradyCCCC05,AymarMP07,KorekIJQC09} \\ 
  & & \cite{DardouriIJQC12,WiatrCP18,FedorovJCP14,ChaiebIJQC14,WiatrPS15} \\
NaCs  
  & \cite{DiemerCPL84,DocenkoEPJD04,DocenkoJPB06} 
  & \cite{IgelJCP86,KorekCJP00,AymarJCP05,DardouriIJQC12,MabroukJPCA14,FedorovJCP14,AymarMP07}\\
  & & \cite{SchwarzerJCP21,KorekJCP07,WeiPRA22} \\
NaFr       & - & \cite{JellaliJPCA18} \\
K$_2$ 
  & \cite{TangoJCP68,EngelkeCPL84,LuhJMS85,RossJPB86,HeinzeJCP87,LiJCP90,AmiotJMS91,ZemkeJCP94,AmiotJCP95,ZhaoJCP96,ZavitsasJCP06,FalkePRA08,PashovEPJD08,LauJCP16,TiemannPRR20}
  & \cite{RoachJMS72,ValancePLA81,MaynauCPL81,ValanceJPB82,PartridgeJCP83,JeungJPB83c,MullerJCP84,IgelJCP86}\\
  & & \cite{JeungJPB88,KraussJCP90,IlyabaevJCP93,MagnierPRA96,LimJCP05,NoroTCA08,JraijJCP09,PershinaCP12} \\
KRb  
  & \cite{RossJPB90,AmiotJCP00,PashovPRA07,StwalleyJCP05,WangPRA07,SchwarzerJCP20}
  & \cite{IgelJCP86,RousseauJMS00,ParkCP00,KotochigovaPRA03,AymarJCP05,SoldanJCP07,StwalleyJPCA10,ChenSAAA12,GuoMP13,FedorovJCP14,ShundalauCTC16,JellaliJQSRT21,JasikADNDT23} \\
KCs  
  & \cite{FerberJCP08,FerberPRA09,FerberPRA13,SchwarzerJCP21,KruminsJCP22,KruminsJQSRT22} 
  & \cite{IgelJCP86,KorekCJP00,AymarJCP05,KotochigovaJCP05,KorekJCP06,KimJMS09,FedorovJCP14,HabliJPB20} \\
KFr  
  & - & - \\
Rb$_2$ 
  & \cite{KotnikCP79,CaldwellCP80,BrefordCPL80,PichlerJPB83,AmiotCPL85,AmiotJCP90,SetoJCP00,BeserJCP09,StraussPRA10,GuanJCP13,LauJCP16,BauerJCP19} 
  & \cite{PartridgeJCP83,SpiegelmannJPB89,IgelJCP86,KraussJCP90,FoucraultJCP92,ParkJMS01,EdvardssonMP03,LimJCP05,AymarJPB06,NoroTCA08} \\
  & & \cite{PershinaCP12,AlloucheJCP12,TomzaPRA12,TomzaMP13,HillJCP17} \\
RbCs  
  & \cite{KatoJCP83,GustavssonCPL88,FellowsJMS99,DocenkoPRA11,TakekoshiPRA12,SchwarzerJCP21} 
  & \cite{IgelJCP86,PavoliniJCP89,AlloucheJPB00,FahsJPB02,AymarJCP05,AymarJPB06,FedorovJCP14,SouissiJQSRT17} \\
RbFr 
  & - & \cite{AymarJPB06,JendoubiTCA16} \\
Cs$_2$ 
  & \cite{KuschJMS69,HoningJCP79,RaabJCP82,RaabCPL82,WeickenmeierJCP85,AmiotJCP88,AmiotJCP02,XieJCP09,CoxonJCP10,XieJCP11,SainisPRA12,SovkovOS13,LauJCP16,SovkovJCP17,BauerJCP19} 
  & \cite{WalchCPL82,JeungJPB83,LaskowskiCPL82,IgelJCP86,MoulletJCP89,KraussJCP90,FoucraultJCP92,LimJCP05,AymarJPB06}\\
  & & \cite{NoroTCA08,AlloucheJCP12,PershinaCP12,HillJCP17} \\
CsFr 
  & - 
  & \cite{AymarJPB06,AlyousefAJBAS20} \\
Fr$_2$ 
  & - 
  & \cite{AymarJPB06,RoosTCA04,LimJCP05,NoroTCA08,PershinaCP12,CamposCPL17,HillJCP17,JellaliJPCA22,NetoJMM22} \\
\end{tabular}
\end{ruledtabular}
\end{table}

\begin{table}[tb!]
\caption{References to previous experimental and theoretical works on the potential energy curves of the alkali-metal--alkaline-earth-metal diatomic molecules in the $X^2\Sigma^+$ electronic state.} 
\label{tab:Ref_AM-AEM}
\begin{ruledtabular}
\begin{tabular}{lll}
Molecule & Experiment & Theory \\
\hline
LiBe & \cite{PersingerJPCA21} & \cite{JonesJCP80,SafonovJSC83,SchlachtaCPL90,FischerCP91,PakJCS91,MarinoJCP92,BauschlicherJCP92,BoldyrevJCP93,KotochigovaJCP11,YouPRA15,PototschnigPCCP16,KoputJCC22} \\
NaBe & - & \cite{BauschlicherJCP92,PototschnigPCCP16,ChmaisaniCP17} \\
KBe  & - & \cite{XiaoJCP13,PototschnigPCCP16,WanCPB17,AwadPS20} \\
RbBe & - & \cite{PototschnigPCCP16,YouJQSRT15,AwadPS20} \\
CsBe & - & \cite{YouJQSRT15,AwadPS20} \\
FrBe & - & - \\
LiMg & \cite{BerryCPL97,PersingerJPCA21b} & \cite{JonesJCP80,FantucciJCP89,BauschlicherJCP92,BoldyrevJCP93,KotochigovaJCP11,GopakumarPRA11,AugustovivcovaJCP12,GaoMP14,PototschnigPCCP16,HouallaCJP19,GaddourJQSRT24} \\
NaMg & - & \cite{BenardCPL82,FantucciJCP89,BauschlicherJCP92,AugustovivcovaJCP12,PototschnigPCCP16,LiSAA18,HouallaCJP19,GaddourJQSRT24} \\
KMg  & - & \cite{AugustovivcovaJCP12,HouallaCTC17,PototschnigPCCP16,GaddourJQSRT24} \\
RbMg & - & \cite{AugustovivcovaJCP12,HouallaCTC17,PototschnigPCCP16} \\
CsMg & - & \cite{AugustovivcovaJCP12,HouallaCTC17} \\
FrMg & - & - \\
LiCa & \cite{WuICMSIP83,RussonJCP98,IvanovaJCP11,KroisJPCA13,GerschmannMP23} &  \cite{NeumannCPL80,JonesJCP80,AlloucheCPL94,RussonJCP98,KotochigovaJCP11,GopakumarPRA11,GopakumarPRA13,GopakumarJCP14,PototschnigPCCP16,PototschnigPRA17} \\
NaCa & - & \cite{GopakumarJCP14,PototschnigPCCP16,PototschnigPRA17,MoussaACSO22} \\
KCa  & \cite{GerschmannPRA17}& \cite{GopakumarJCP14,PototschnigPCCP16,PototschnigPRA17,MoussaNJP21} \\
RbCa & - & \cite{GopakumarJCP14,PototschnigPCCP16,PototschnigJMS15,PototschnigPRA17} \\
CsCa & - & \cite{MoussaACSO22} \\
FrCa & - & -\\
LiSr & \cite{SchwankeJPB17} & \cite{GueroutPRA10,GopakumarPRA11,KotochigovaJCP11,GopakumarPRA13,GuoMP13,GopakumarJCP14,PototschnigPCCP16,PototschnigPRA17,ZeidCTC18} \\
NaSr & - & \cite{GueroutPRA10,GopakumarJCP14,PototschnigPCCP16,PototschnigPRA17,ZeidCTC18} \\
KSr  & \cite{SzczepkowskiJQSRT18}& \cite{GueroutPRA10,GopakumarJCP14,PototschnigPCCP16,PototschnigPRA17,SzczepkowskiJQSRT18,ZeidCTC18} \\
RbSr & \cite{PototschnigJCP14,KroisPCCP14,LacknerPRL14,CiameiPCCP18} & \cite{ZuchowskiPRL10,GueroutPRA10,ChenPRA14,PototschnigJCP14,ZuchowskiPRA14,PototschnigPCCP16,GopakumarJCP14,PototschnigPRA17} \\
CsSr & - & \cite{GueroutPRA10} \\
FrSr & - & - \\
LiBa & \cite{StringatJMS94}& \cite{AlloucheJCP94,GopakumarPRA11,GouJCP15} \\
NaBa & - & \cite{BoutassettaCP94,GouJCP15} \\
KBa  & - & \cite{BoutassettaCP95,GouJCP15} \\
RbBa & - & \cite{GouJCP15,LadjimiJQSRT20} \\
CsBa & - & \cite{GouJCP15,LadjimiTCA19} \\
FrBa & - & - \\
LiRa & - & \cite{FleigNJP21} \\
NaRa & - & \cite{FleigNJP21} \\
KRa  & - & \cite{FleigNJP21} \\
RbRa & - & \cite{FleigNJP21} \\
CsRa & - & \cite{FleigNJP21} \\
FrRa & - & \cite{FleigNJP21} \\
\end{tabular}
\end{ruledtabular}
\end{table}

\begin{table}[tb!]
\caption{References to previous experimental and theoretical works on the potential energy curves of the alkaline-earth-metal diatomic molecules in the $X^1\Sigma^+$ electronic state.} 
\label{tab:Ref_AEM-AEM}
\begin{ruledtabular}
\begin{tabular}{lll}
Molecule & Experiment & Theory \\
\hline
Be$_2$ & \cite{BondybeyCPL84,SchautzTCA98,KaledinJMS99,SpirkoJMS06,MerrittScience09,MeshkovJCP14} &  \cite{LiuJCP80,ChilesJCP81,KaledinJMS99,RoeggenIJQC05,ZhaoJPCA06,PatkowskiJPCA07,PatkowskiScience09,SchmidtJPCA10, 
      HeavenCPL11,KoputPCCP11,LiJPCA11,JeungCRC12,HelalCPL13,ShengPRA13,SharmaJCP14,KhatibJPCA14,LesiukPRA15,DeibleJCP15,KalemosJCP16,   
      MagoulasJPCA18,LesiukJCTC19,BaoJPCL19,ChengIJQC19,XuJCP20,DerbovJQSRT21,GutherJCP21}\\
BeMg   & - & \cite{ChilesCPL82,BorinCPL03,RuetteJMS05,HeavenCPL11,KerkinesJCP12,MahapatraMP15} \\
BeCa   & - & \cite{HeavenCPL11} \\
BeSr   & - & \cite{HeavenCPL11} \\
BeBa   & - & \cite{HeavenCPL11} \\
BeRa   & - & - \\
Mg$_2$ & \cite{BalfourCJP70,VidalJMS77,WuICMSIP83,KnockelJCP13} & \cite{ChilesJCP81,PartridgeJCP90,DyallJCP92,CzuchajTCA01,TiesingaPRA02,ZhaoJPCA06,PatkowskiJPCA07,LiJCP10,HeavenCPL11,LiJPCA11,AmaranJCP13,YuwonoMP19}\\
 &  &  \cite{BaoJPCL19,ChengIJQC19,YuwonoSA20,DyallJCTC23} \\
MgCa   & \cite{AtmanspacherJCP85} & \cite{HeavenCPL11,WeiCPL15,RizkallahCP21} \\
MgSr   & - & \cite{HeavenCPL11,WeiCPL15} \\
MgBa   & - & \cite{HeavenCPL11,WeiCPL15} \\
MgRa   & - & - \\
Ca$_2$ & \cite{BalfourCJP75,WyssJCP79,VidalJCP80,WuICMSIP83,AllardPRA02,AllardEPJD03} &  \cite{DyallJCP92,CzuchajCPL03,BusseryPRA03,ZhaoJPCA06,PatkowskiJPCA07,RoyMP07,YangJCP09,BouissouJCP10,HeavenCPL11,LiJPCA11,YangTCA12}\\
 & & \cite{MosyaginIJQC13,BaoJPCL19,ChengIJQC19} \\
CaSr   & - & \cite{HeavenCPL11,WeiCPL15} \\
CaBa   & - & \cite{HeavenCPL11,WeiCPL15} \\
CaRa   & - & - \\
Sr$_2$ & \cite{BergemanJCP80,GerberJCP84,SteinPRA08,SteinEPJD10,LeungNJP21} &  \cite{BoutassettaPRA96,WangJPCA00,CzuchajCPL03b,KotochigovaJCP08,MitinRJPCA09,YinJCP10,LiJPCA11,HeavenCPL11,SkomorowskiJCP12,YangTCA12,BaoJPCL19,ChengIJQC19,Szczepkowski2023,DyallJCTC23} \\
SrBa   & - & \cite{HeavenCPL11,WeiCPL15} \\
SrRa   & - & -\\
Ba$_2$ &\cite{LebeaultJMS98} & \cite{AlloucheCP95,MitinRJPCA09,LiJPCA11,HeavenCPL11,YangTCA12,BaoJPCL19,ChengIJQC19} \\
BaRa   & - & - \\
Ra$_2$ & - & \cite{RoosTCA04,TeodoroJCP15,CamposCPL17,BaoJPCL19,NetoJMM22} \\
\end{tabular}
\end{ruledtabular}
\end{table}

In this work, we use state-of-the-art \textit{ab initio} electronic structure methods to calculate the potential energy curves, permanent electric dipole moments, and static electric dipole polarizabilities for all the hetero- and homonuclear diatomic molecules composed of alkali-metal (Li, Na, K, Rb, Cs, Fr) and alkaline-earth-metal (Be, Mg, Ca, Sr, Ba, Ra) atoms. We employ the hierarchy of coupled cluster methods upto CCSDTQ with large Gaussian basis sets and small-core relativistic energy-consistent pseudopotentials. We study a large number of 78 hetero- and homonuclear diatomic combinations, including 21 alkali-metal molecules in the $X^1\Sigma^+$ and $a^3\Sigma^+$ electronic states, 36 alkali-metal--alkaline-earth-metal molecules in the $X^2\Sigma^+$ electronic state, and 21 alkaline-earth-metal molecules in the $X^1\Sigma^+$ electronic state. We also analyze the convergence and accuracy of our calculations with the size of the orbital basis sets and the quality of the wave functions. In this way, we establish a new theoretical benchmark. 

A significant portion of the molecules under investigation has already been studied experimentally or theoretically. Notably, alkali-metal dimers are one of the most extensively studied classes of molecules. However, alkali-metal--alkaline-earth-metal and alkaline-earth-metal molecules have received less attention, particularly in experimental studies, with a large number of such combinations without any experimental data. In the following sections, we will compare our results with the most recent and accurate experimental and theoretical values. Because of the large number of studied molecules, refereeing and comparing to all previous theoretical results is not feasible. Therefore, we have collected references to previous experimental and theoretical works on the potential energy curves of the alkali-metal--alkali-metal, alkali-metal--alkaline-earth-metal, and alkaline-earth-metal--alkaline-earth-metal diatomic molecules in their ground electronic states in Tables~\ref{tab:Ref_AM-AM},~\ref{tab:Ref_AM-AEM}, and~\ref{tab:Ref_AEM-AEM}, respectively.

Experimental works in this field can be categorized into several classes, including: 1) laser photoionization spectroscopy, 2) laser-induced fluorescence spectroscopy, 3) polarisation labeling spectroscopy in hot vapors or beams, 4) spectroscopy of molecules on helium nanodroplets, and 5) highly accurate measurements in ultracold atomic or molecular gases in traps. Previous theoretical works can also be classified into several groups, including: 1) the oldest results, often without including the electron correlation, 2) the results with large-core pseudopotentials and the full conﬁguration interaction method for valence electrons, 3) the results with small-core (scalar-relativistic) pseudopotentials and truncated coupled cluster or multireference configuration interactions methods, and 4) all-electron calculations often with relativistic Hamiltonians.

The main advancements of the present work lie in the following:
\begin{itemize}
\item the coherent calculation and comparison of three classes of experimentally relevant molecules at the consistent level of theory,
\item the computation of electronic properties of some molecules, such as those containing francium and radium, for the first time,
\item the investigation of permanent electric dipole moments of heteronuclear alkaline-earth molecules, for the first time,
\item the computation of electronic properties of some molecules, such as those containing francium and radium, for the first time,
\item the use of large recently developed Gaussian basis sets~\cite{HillJCP17},
\item the inclusion of full triple and quadruple excitations in the coupled cluster method to obtain potential energy curves, resulting in the description of valence electrons at the full configuration interaction level for all molecules,
\item the publication of all results in a numerical form in the Supplemental Material~\cite{supplemental}.
\end{itemize}

For completeness, it is worth mentioning the existence of the studies of isoelectronic diatomic molecules containing: 1) alkaline-earth-metal-like ytterbium atom~\cite{BuchachenkoEPJD07,KitagawaPRA08,TecmerIJQC19}, including its combinations with alkali-metal~\cite{SorensenJPCA09,ZhangJCP10,GopakumarJCP10,MunchowPCCP11,TohmeCP13,BorkowskiPRA13,TohmeJCP15,TohmeCPL15,TohmeCTC16,ShaoJPCA17,MeniailavaCTC17,ShundalauCTC17,GuttridgePRA18,GreenPRA19} or alkaline-earth-metal~\cite{TomzaPCCP11,ShaoJPCA17} atom, 2) alkaline-earth-metal-like zinc, cadmium, or mercury atom with alkali-metal or alkaline-earth-metal atom~\cite{BorkowskiPRA17,ZarembaPRA21}, and 3) alkali-metal-like copper, silver, or gold atom with alkali-metal or alkaline-earth-metal atom~\cite{SmialkowskiPRA21,KlosNJP22,XiaoCPL22}. Finally, diatomic molecular ions of alkali-metal and alkaline-earth-metal atoms have also been investigated extensively (see, e.g.,~Ref.~\cite{SmialkowskiPRA20} and references therein). Such systems are, however, out of the scope of this investigation.
 
The structure of the paper is the following. In Section~\ref{sec:methods}, we describe the employed computational methods. In Section~\ref{sec:results}, we present and discuss the obtained results. In section~\ref{sec:summary}, we provide a summary and outlook.

\section{COMPUTATIONAL METHODS}
\label{sec:methods} 

We adopt and employ the computational scheme based on the composite approach to calculate potential energy curves in the Born-Oppenheimer approximation for the relevant molecular electronic states, which we established and tested in several previous studies on different classes of molecules containing alkali-metal and alkaline-earth-metal atoms~\cite{GronowskiPRA20,SmialkowskiPRA20,SmialkowskiPRA21,ZarembaPRA21}. The final interaction energies, $V_\text{int}(R)$, as a function of the internuclear distance $R$, are calculated as a sum of different contributions
\begin{equation}\label{eq:Vint}
\begin{split}
V_\text{int}(R)=& V_\text{CCSD(T)}^\text{apwCV5Z+bf}(R)+\delta V_\text{CCSDT}^\text{apwCVTZ}(R)\\
                &+\delta V_\text{CCSDTQ,val}^\text{apVTZ}(R)\,,
\end{split}
\end{equation}
where the leading part, $V_\text{CCSD(T)}^\text{apwCV5Z+bf}(R)$, is obtained with the closed-shell (for the $X^1\Sigma^+$ states) or spin-restricted open-shell (for the $X^2\Sigma^+$ and  $a^3\Sigma^+$ states) coupled cluster method restricted to single, double, and noniterative triple excitations [CCSD(T)]~\cite{MusialRMP07} and the augmented correlation-consistent polarized weighted core-valence quintuple-$\zeta$ quality basis sets (aug-cc-pwCV5Z)~\cite{PrascherTCA10,HillJCP17}. The atomic basis sets are additionally augmented in these calculations by the set of the $[4s4p3d3f1g]$ bond functions (bf)~\footnote{Bond function exponents, $s$: 0.6, 0.2, 0.067, 0.02, $p$: 0.6, 0.2, 0.067, 0.02, $d$: 0.4, 0.13, 0.04, $f$: 0.4, 0.13, 0.04, $g$: 0.2.} to accelerate the convergence toward the complete basis set limit~\cite{midbond}. 

The next leading electron-correlation correction to the interaction energy, to account for the contribution of the iterative full triple excitations in the coupled cluster method, $\delta V_\text{CCSDT}^\text{apwCVTZ}(R)$, is obtained as  
\begin{equation}\label{eq:VSDT}
\delta V_\text{CCSDT}^\text{apwCVTZ}(R)=V_\text{CCSDT}^\text{apwCVTZ}(R)-V_\text{CCSD(T)}^\text{apwCVTZ}(R)\,,
\end{equation}
where $V_\text{CCSDT}^\text{apwCVTZ}(R)$ is the interaction energy calculated with the coupled cluster method restricted to single, double, and triple excitations (CCSDT) and $V_\text{CCSD(T)}^\text{apwCVTZ}(R)$ -- with the CCSD(T) method, both with the same augmented correlation-consistent polarized weighted core-valence triple-$\zeta$ quality basis sets (aug-cc-pwCVTZ)~\cite{PrascherTCA10,HillJCP17}.

Finally, the valence electron-correlation correction to the interaction energy to account for the contribution of the quadruple excitations in the coupled cluster method, $\delta V_\text{CCSDTQ,val}^\text{apVTZ}(R)$, is obtained as 
\begin{equation}\label{eq:VSDTQ}
\delta V_\text{CCSDTQ,val}^\text{apVTZ}(R)=V_\text{CCSDTQ,val}^\text{apVTZ}(R)-V_\text{CCSDT,val}^\text{apCVTZ}(R)\,,
\end{equation}
where $V_\text{CCSDTQ,val}^\text{apVTZ}(R)$ is the interaction energy calculated with the coupled cluster method restricted to single, double, triple, and quadruple excitations (CCSDTQ) and $V_\text{CCSDT,val}^\text{apCVTZ}(R)$ -- with the CCSDT method, both with the same augmented correlation-consistent polarized valence triple-$\zeta$ quality basis sets (aug-cc-pVTZ)~\cite{PrascherTCA10,HillJCP17}. In these calculations, only valence electrons are correlated, therefore $\delta V_\text{CCSDTQ,val}^\text{apVTZ}(R)$ is identically equal to zero for alkali-metal and alkali-metal--alkaline-earth-metal molecules with two and three valence electrons, respectively. 

The interaction energies, $V_{method}^{basis}(R)$, in Eqs.~\eqref{eq:Vint}-\eqref{eq:VSDTQ} are obtained using the super-molecule approach with the basis set superposition error (BSSE) corrected by using the Boys-Bernardi counterpoise correction~\cite{BoysMP70},
\begin{equation}
V_{method}^{basis}(R)=E_{AB}(R)-E_A(R)-E_B(R)\,,
\end{equation}
where $E_{AB}(R)$ is the total energy of the molecule $AB$, and $E_{A}(R)$ and $E_{B}(R)$ are the total energies of the atoms $A$ and $B$, all computed with the given $method$ and diatom $basis$ set at a distance $R$.

The employed computational scheme given by Eq.~\eqref{eq:Vint} works very well at all internuclear distances for all molecules in the considered doublet $X^2\Sigma^+$ and triplet $a^3\Sigma^+$ molecular electronic states and for alkaline-earth-metal molecules in the $X^1\Sigma^+$ states because they are well described at all internuclear distances by single-reference methods. On the other hand, the singlet $X^1\Sigma^+$ molecular electronic states of the alkali-metal molecules have single-reference nature at small and intermediate internuclear distances and multireference nature at larger distances, which originates from the open-shell character of the interacting alkali-metal atoms. At larger distances, the single-reference CCSD(T) method gives incorrect results, and the CCSDT correction only partially improves them. Therefore, we compute the $X^1\Sigma^+$ states of alkali-metal molecules with the coupled cluster methods in the vicinity of the potential well at short and intermediate distances and smoothly merge them at larger distances with the long-range multiple expansion of the interaction energy asymptotically given by
\begin{equation}
V_\text{int}(R)\approx -\frac{C_6}{R^6}-\frac{C_8}{R^8}-\frac{C_{10}}{R^{10}}\,,
\end{equation}
where the three leading long-range dispersion coefficients $C_6$, $C_8$, and $C_{10}$ are taken from accurate atomic calculations reported in Ref.~\cite{JiangADNDT15} (for molecules containing Fr, only $C_6$ from Ref.~\cite{DereviankoPRA01} are used). Depending on the system, the switching distance is around 11-14$\,$bohr.

All electrons in the Li, Be, Na, and Mg atoms and outer-shell electrons in other atoms are explicitly described by the selected large atomic augmented correlation-consistent one-electron Gaussian basis sets, while the inner-shell electrons in heavier atoms are replaced by the small-core relativistic energy-consistent pseudopotentials (ECP)~\cite{DolgCR12} to include the scalar relativistic effects. The ECP10MDF, ECP10MDF, ECP28MDF, ECP28MDF, ECP46MDF, ECP46MDF, ECP78MDF, and ECP78MDF pseudopotentials~\cite{LimJCP05,LimJCP06} are used for the K, Ca, Rb, Sr, Cs, Ba, Fr, and Ra atoms, respectively. The electrons of two outermost shells, i.e., $(n-1)s^2(n-1)p^6{n}s^1$ from alkali-metal and $(n-1)s^2(n-1)p^6{n}s^2$ from alkaline-earth-metal atoms, are correlated in calculations with the aug-cc-pwCV$n$Z basis sets, while only valence electrons, i.e., ${n}s^1$ from alkali-metal and ${n}s^2$ from alkaline-earth-metal atoms -- with the aug-cc-pV$n$Z basis sets.

The composite approach of Eq.~\eqref{eq:Vint} rely on error cancellation observed in high-level molecular electronic structure calculations~\cite{LesiukJCTC19,GronowskiPRA20,LesiukPRA20,GebalaPRA23}. Effectively, for three classes of molecules, we describe their valence electrons at the full configuration interaction (FCI) level because CCSD, CCSDT, and CCSDTQ are equivalent to FCI for two-, three-, and four-electron systems, respectively. In order to evaluate the accuracy of the used approach, additional test calculations are carried out for exemplary KRb, RbSr, and CaSr molecules. Different basis sets from the aug-cc-pwCV$n$Z and aug-cc-pV$n$Z families with $n$=D, T, Q, 5 are employed and extrapolate the complete basis set limit (CBS) using the two-point formula~\cite{HelgakerJCP97}. Additionally, potential energy curves are obtained with a series of different less accurate methods~\cite{Helgaker}: spin-restricted Hartree-Fock (RHF), second-order M{\o}ller-Plesset perturbation theory (MP2), coupled cluster method restricted to single and double excitations (CCSD), configuration interaction method restricted to single and double excitations (CISD) and its variant including the Davidson correction (CISD+Q), as well as multireference variants of the CISD and CISD+Q methods (MRCISD and MRCISD+Q), respectively. 

We interpolate the potential energy curves, $V_\text{int}(R)$, using the cubic spline method to obtain spectroscopic parameters. The equilibrium distance, $R_e$, is defined by
\begin{equation}
\left.\frac{dV_\text{int}(R)}{dR}\right|_{R_e}=0\,,
\end{equation}
and the corresponding potential energy well depth, $D_e$, is given by
\begin{equation}
D_e = -V_\text{int}(R_e)\,.
\end{equation}
The harmonic constant, $\omega_e$, of the interaction potential is calculated at the equilibrium distance as
\begin{equation}
\omega_e = \sqrt{\frac{1}{\mu}\left.\frac{d^2V_\text{int}(R)}{dR^2}\right|_{R_e}}\,,
\end{equation}
where $\mu$ represents the reduced mass. The equilibrium rotational constant, $B_e$, is given as
\begin{equation}
B_e = \frac{\hbar^2}{2 \mu R_e}\,.
\end{equation}
Finally, the first anharmonicity constant, $\omega_ex_e\approx -Y_{20}$, is obtained by fitting the Dunham expansion to the lowest vibrational energy levels obtained with the discrete variable representation (DVR).

The permanent electric dipole moments $d(R)$ and static electric dipole polarizabilities $\alpha(R)$ are calculated with the finite field approach using the CCSD(T) method and the aug-cc-pwCV5Z basis sets with bond functions. The $z$ axis is chosen along the internuclear axis and oriented from more electronegative atoms to less electronegative ones. The value of the external field perturbation equal to $\pm$0.0005$\,$a.u. is used.

All electronic structure calculations are performed with the \textsc{Molpro}~\cite{Molpro,MOLPRO-WIREs} and \textsc{MRCC}~\cite{KallayJCP20} packages of \textit{ab initio} programs. Atomic masses of the most abundant isotopes are assumed.

\section{RESULTS AND DISCUSSION}
\label{sec:results}

\begin{figure*}[tb]
\begin{center}
\includegraphics[width=\textwidth]{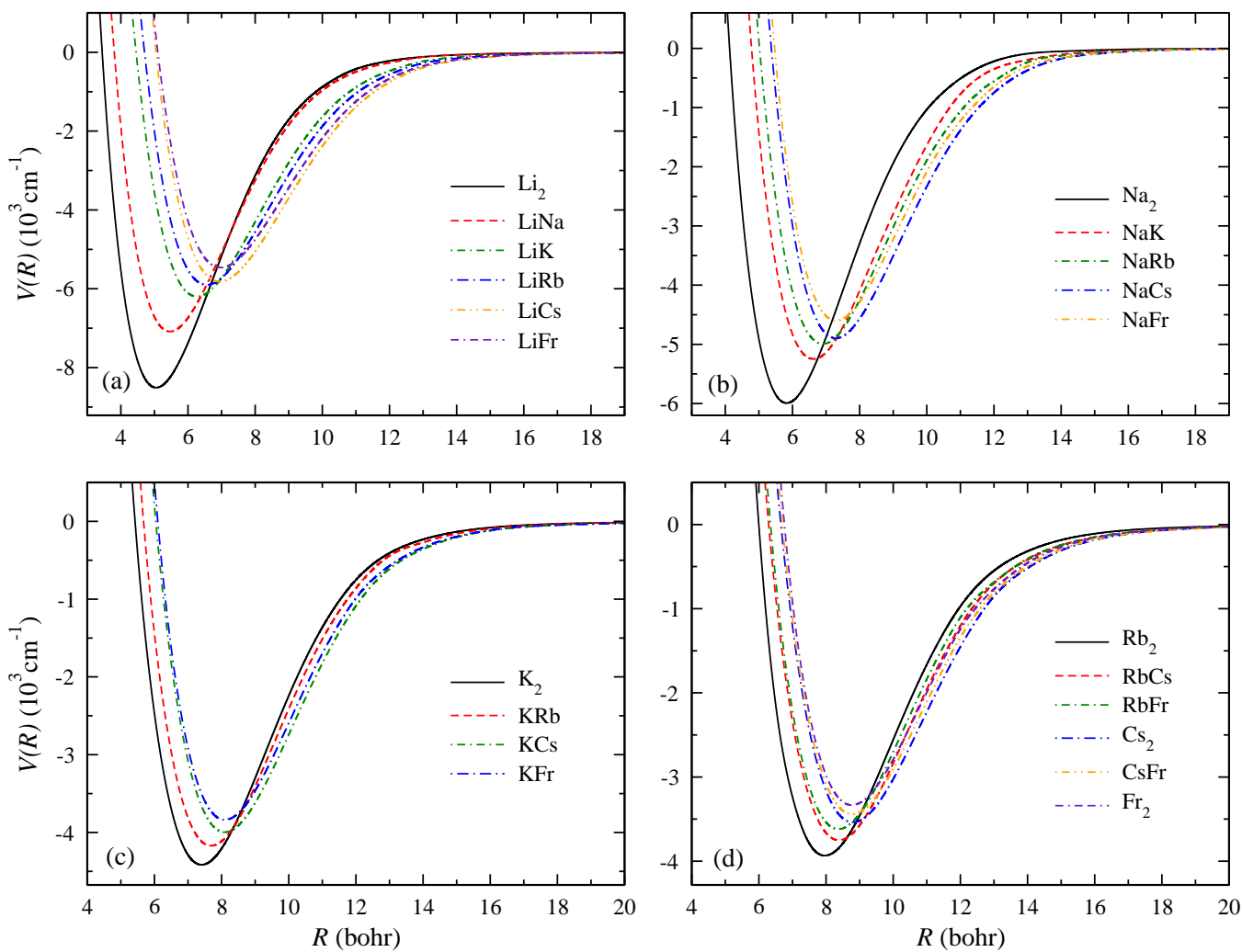}
\caption{Potential energy curves of all the alkali-metal diatomic molecules in the ground $X^1\Sigma^+$ electronic state.}
\label{fig:AM-AM-S}
\end{center}
\end{figure*}

\begin{figure*}[tb]
\begin{center}
\includegraphics[width=\textwidth]{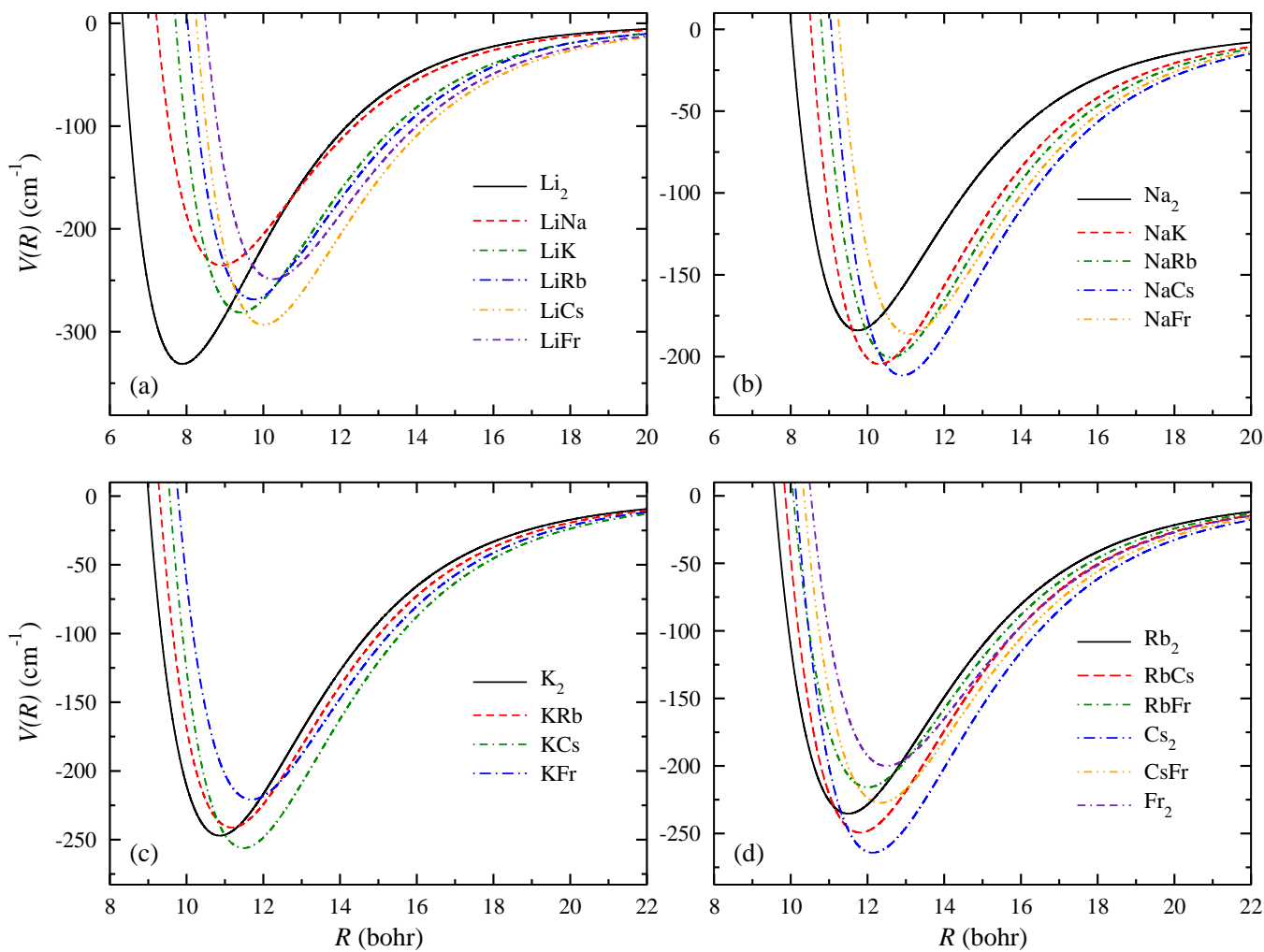}
\caption{Potential energy curves of all the alkali-metal diatomic molecules in the $a^3\Sigma^+$ electronic state.}
\label{fig:AM-AM-T}
\end{center}
\end{figure*}

\begin{figure*}[!tb]
    \begin{center}
    \includegraphics[width=\textwidth]{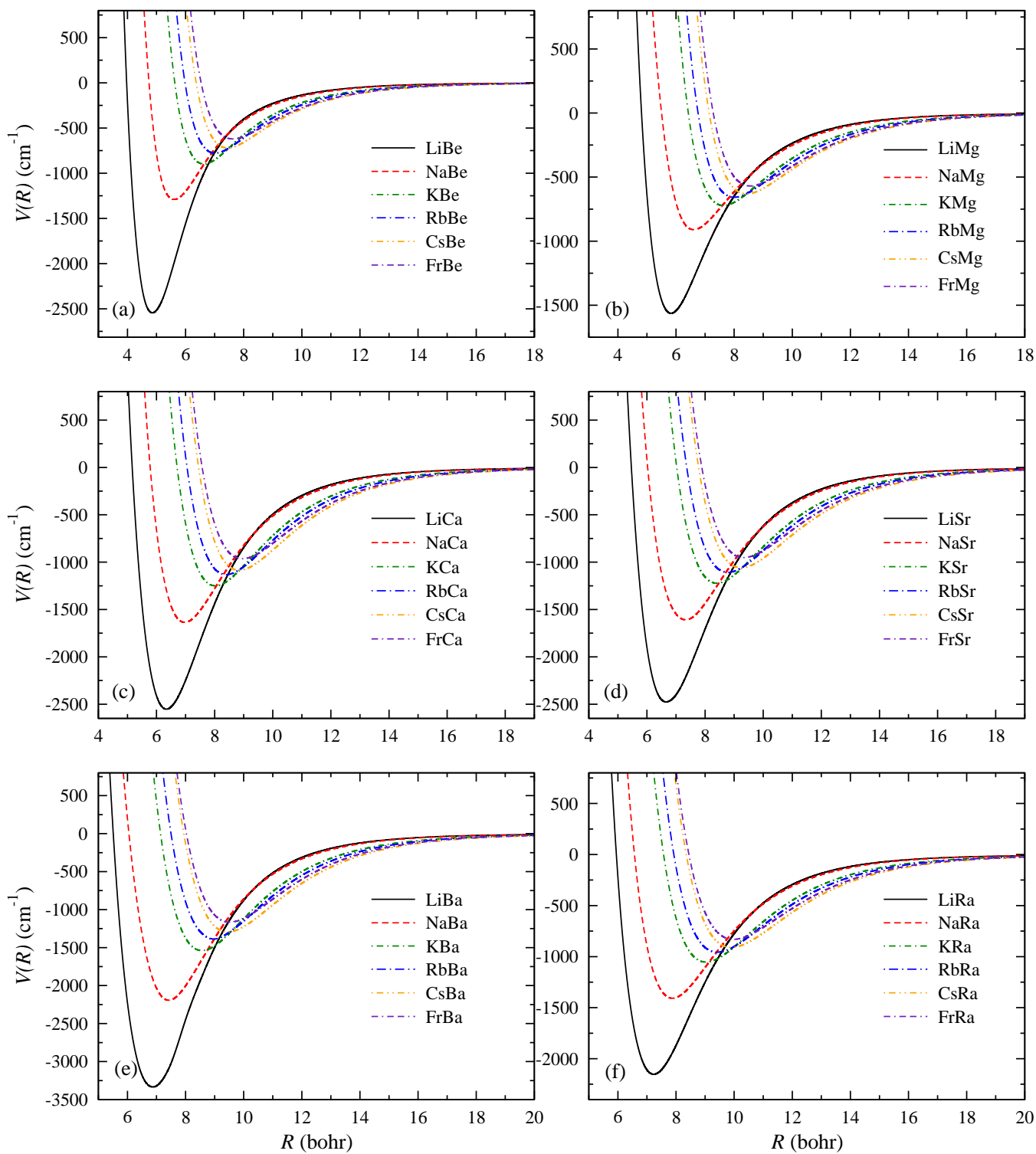}
    \caption{Potential energy curves of all the alkali-metal--alkaline-earth-metal diatomic molecules in the ground $X^2\Sigma^+$ electronic state.}
    \label{fig:AM-AEM}
    \end{center}
\end{figure*}

\begin{figure*}[!htbp]
    \begin{center}
    \includegraphics[width=\textwidth]{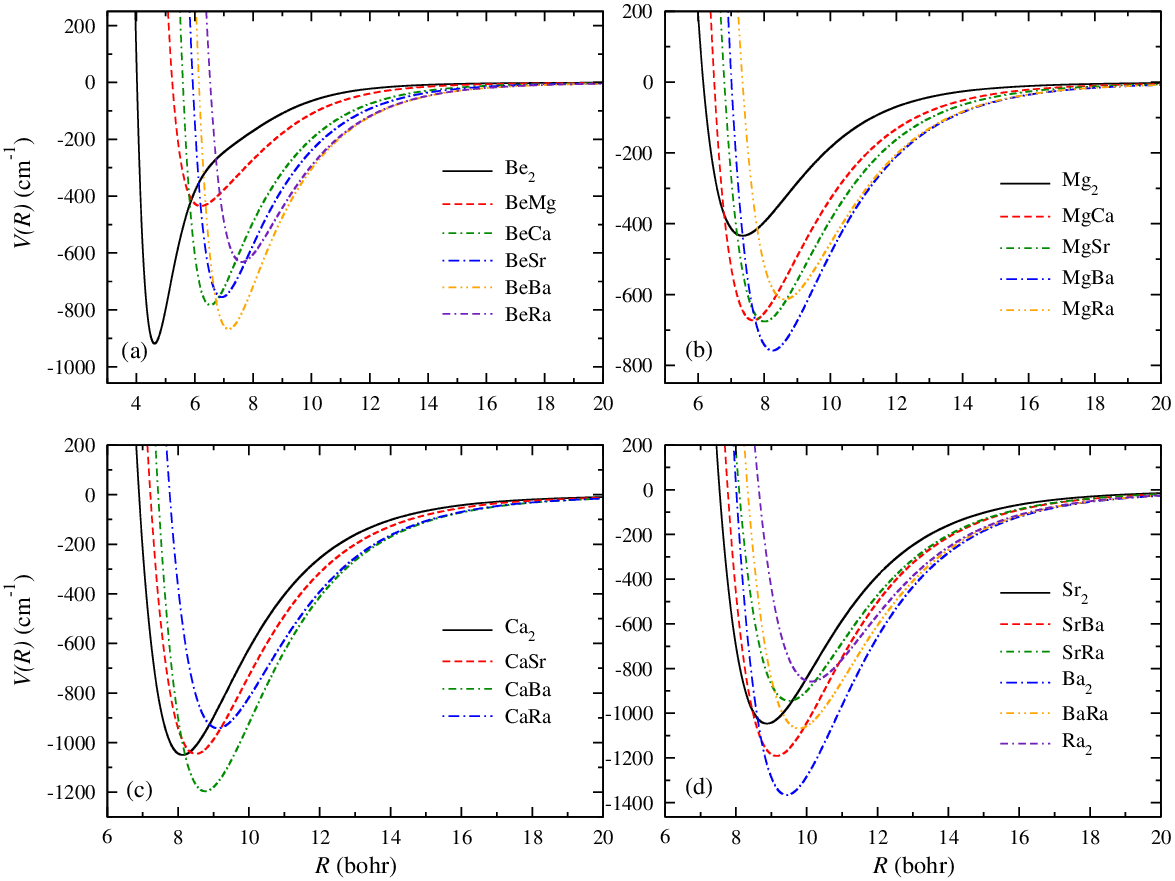} 
    \caption{Potential energy curves of all the alkaline-earth-metal diatomic molecules in the ground $X^1\Sigma^+$ electronic state.} 
    \label{fig:AEM-AEM}
    \end{center}
\end{figure*}

The interaction of two open-shell alkali-metal atoms in their ground doublet $^2S$ electronic state results in the ground molecular electronic state of the singlet $X^1\Sigma^+$ symmetry and the first excited electronic state of the triplet $a^3\Sigma^+$ symmetry. The interaction of an alkali-metal atom with a closed-shell alkaline-earth-metal atom results in the ground molecular electronic state of the doublet $X^2\Sigma^+$ symmetry. Finally, the interaction of two alkaline-earth-metal atoms in their ground single $^1S$ electronic state results in the ground molecular electronic state of the singlet $X^1\Sigma^+$ symmetry. Additionally, homonuclear dimers have the gerade and ungerade symmetries in the singlet ($X^1\Sigma^+_g$) and triplet ($a^3\Sigma^+_u$) states, respectively.

\subsection{Potential energy curves}

The computed potential energy curves for the alkali-metal diatomic molecules in the $X^1\Sigma^+$ and $a^3\Sigma^+$ electronic states are presented in Fig.~\ref{fig:AM-AM-S} and Fig.~\ref{fig:AM-AM-T}, respectively, for the alkali-metal--alkaline-earth-metal diatomic molecules in the $X^2\Sigma^+$ electronic state -- in Fig.~\ref{fig:AM-AEM}, and for the alkaline-earth-metal diatomic molecules in the $X^1\Sigma^+$ state -- in Fig.~\ref{fig:AEM-AEM}. Calculations are performed for all the combinations of the alkali-metal (Li, Na, K, Rb, Cs, Fr) and alkaline-earth-metal (Be, Mg, Ca, Sr, Ba, Ra) atoms. All potential energy curves show a smooth behavior with well-defined minima. The corresponding spectroscopic characteristics such as the equilibrium interatomic distance~$R_e$, well depth~$D_e$, harmonic constant~$\omega_e$, first anharmonicity constant~$\omega_ex_e$, and rotational constant~$B_e$ are collected in Tables~\ref{table:AM-AM-S}-\ref{table:AEM-AEM}, along with available experimental spectroscopic data.

\subsubsection{Alkali-metal diatomic molecules}

The alkali-metal diatomic molecules in their ground $X^1\Sigma^+$ electronic state are relatively strongly chemically bound. Formally, they are covalently bound with a bond order of one. The calculated well depths systematically decrease with increasing the atomic number of the alkali-metal atoms and range from 3333$\,$cm$^{-1}$ for Fr$_2$ to 8509$\,$cm$^{-1}$ for Li$_2$ (with an average of 3892$\,$cm$^{-1}$ calculated over a set of all molecules of their type). The corresponding equilibrium distances systematically increase with increasing the atomic number of the alkali-metal atoms and take values between 5.05$\,$bohr for Li$_2$ and 8.80$\,$bohr for Cs$_2$ (with an average of 7.31$\,$bohr). 

The alkali-metal diatomic molecules in their first excited $a^3\Sigma^+$ electronic state with two valence electrons spin-polarized are weakly bound van der Waals complexes. Formally, they are not chemically bound with a bond order of zero and are stabilized by the dispersion interaction. The calculated well depths range from 198$\,$cm$^{-1}$ for Fr$_2$ to 334$\,$cm$^{-1}$ for Li$_2$ (with an average of 239$\,$cm$^{-1}$), displaying less variation and regularity than moleucles in their ground $X^1\Sigma^+$ states. The corresponding equilibrium distances systematically increase with increasing the atomic number of the alkali-metal atoms and take values between 7.88$\,$bohr for Li$_2$ and 12.51$\,$bohr for Fr$_2$ (with an average of 10.78$\,$bohr). Thus, the alkali-metal diatomic molecules in the $a^3\Sigma^+$ state have well depths more than an order of magnitude smaller and equilibrium distances almost twice as large as those in their $X^1\Sigma^+$ states.

All the alkali-metal diatomic molecules in the $X^1\Sigma^+$ and $a^3\Sigma^+$ electronic states consisting of stable isotopes (15 atomic combinations) have already been studied experimentally (see Table~\ref{tab:Ref_AM-AM}). Experimental high-resolution spectroscopic studies typically provide potential energy curves with accuracy better than 1$\,$cm$^{-1}$. To evaluate the accuracy of our results, we compare our spectroscopic constants
with the available experimental values in Tables~\ref{table:AM-AM-S} and \ref{table:AM-AM-T} and calculate the root-mean-square error (RMSE) between accurate experimental and present well depths and equilibrium distances.

\begin{table*}[tbp]
\begin{ruledtabular}
\caption{Spectroscopic constants of the alkali-metal diatomic molecules in the ground $X^1\Sigma^+$ electronic state: equilibrium interatomic distance $R_e$, well depth $D_e$, harmonic constant $\omega_e$, first anharmonicity constant $\omega_ex_e$, rotational constant $B_e$, permanent electric dipole moments $d_e$, and parallel $\alpha_e^\parallel$ and perpendicular $\alpha_e^\perp$ components of the static electric dipole polarizability. Available experimental values are also collected.} 
\begin{tabular*}{\textwidth}{lrrrrrrrrr}
Molecule & $R_e\,$(bohr) & $D_e\,$(cm${}^{-1}$) & $\omega_e\,$(cm${}^{-1}$) &$\omega_e\chi_e\,$(cm${}^{-1}$) & $B_e\,$(cm${}^{-1}$) & $d_e\,$(D) & $\alpha_e^\parallel\,$(a.u.) & $\alpha_e^\perp\,$(a.u.) & Ref. \\
 \hline
Li$_2$ 
  &  5.053       & 8509        & 351.2 & 2.59 & 0.6722 & 0 & 296.2 & 165.7 & This work\\
  &  5.051225(4) & 8516.774(4) & 351.4 & 2.58 & 0.6726 & 0 & & & \cite{LeRoyJCP09,HesselJCP79} \\
LiNa 
  &  5.464   & 7085        & 256.4  & 1.69 & 0.3751 & 0.47     & 342.3 & 185.6 & This work\\
  &  5.45859 & 7103.5(1.0) & 256.46 & 1.58 & 0.3758 & 0.47(3) &        &        & \cite{DagdigianJCP71,FellowsJCP91,SteinkePRA12}\\
LiK 
  &  6.269     & 6197      & 211.9     & 1.21     & 0.2576    & 3.36    & 480.3 & 245. 5 & This work\\ 
  &  6.2722(1) & 6216.9(1) & 211.91(2) & 1.224(6) & 0.2576(2) & 3.45(1) &       &        &\cite{TiemannPRA09,EngelkeCP84,DagdigianJCP72}\\
LiRb 
  & 6.558    & 5902    &  195.0  & 1.14  & 0.2160 & 4.00    & 531.5 & 261.6 & This work\\
  & 6.5501(1)& 5928(4) &  195.18 & 1.096 & 0.2165 & 4.00(1) & & & \cite{IvanovaJCP11a,DagdigianJCP72} \\
LiCs 
  & 6.943  &  5817      & 184.4  & 1.05  & 0.1874 & 5.28   & 605.7 & 284.6 & This work\\
  & 6.9317 &  5875.5(1) & 184.70 & 1.000 & 0.1880 & 5.5(2) &       &        & \cite{StaanumPRA07,DeiglmayrPRL08} \\
LiFr 
  & 6.976 & 5460 & 180.7 & 1.79 & 0.1819 & 4.24 & 605.1 & 276.5 & This work\\
Na$_2$ 
  & 5.826  & 5999        & 159.5  & 0.90  & 0.1543 & 0 & 378.8 & 204.7 & This work\\
  & 5.8191 & 6022.029(5) & 159.18 & 0.760 & 0.1547 & 0 &       &       & \cite{BarrowCPL84,JonesPRA96} \\
NaK 
  & 6.629  & 5248       & 120.2  & 0.20  & 0.0947 & 2.71    & 521.8 & 274.7 & This work\\
  & 6.6122 & 5273.6(1)  & 124.03 & 0.496 & 0.0952 & 2.72(6) &       &        & \cite{DagdigianJCP72,GerdesEPJD08,GerdesEPJD11}\\
NaRb  
  & 6.902  & 4993       & 105.7 & 0.16  & 0.0700  & 3.30   & 574.3 & 293.4 & This work\\
  & 6.8850 & 5030.50(5) & 106.85 & 0.380 & 0.0702  & 3.2(1) &       &       & \cite{DocenkoPRA04,PashovPRA05,GuoPRL16} \\
NaCs 
  & 7.295  & 4897       & 100.0 & 0.39  & 0.0577 & 4.52   & 665.9 & 323.7 & This work\\
  & 7.2766 & 4954.24(1) & 98.88 & 0.321 & 0.0580 & 4.8(2) &       &       & \cite{DocenkoEPJD04,DocenkoJPB06,DagdigianJCP72}\\
NaFr 
  & 7.301 & 4599 & 95.8 & 0.20 & 0.0542  & 3.51 & 646.7 & 309.7 & This work\\
K$_2$ 
  & 7.405    & 4415       &   92.6    & 0.32  & 0.0564 & 0 & 705.7 & 372.0 & This work\\
  & 7.415(2) & 4450.67(7) & 92.40 & 0.324 & 0.0562 & 0 &       &       & \cite{AmiotJCP95,ZhaoJCP96}\\
KRb  
  & 7.687     & 4171        & 76.2  & 0.28  & 0.0382 & 0.63    & 762.4 & 398.6 & This work\\
  & 7.6868(1) & 4217.82(1) & 75.85 & 0.230 & 0.0381 & 0.57(2) &       &       & \cite{AmiotJCP00,PashovPRA07,NiScience08} \\
KCs
  & 8.097     & 3996        & 68.5      & 0.20     & 0.0305     & 1.86 & 879.0 & 445.2 & This work\\
  & 8.096(13) & 4069.3(1.5) & 68.394(3) & 0.193(1) & 0.03048(5) &      &       &      & \cite{FerberJCP08} \\
KFr 
  & 8.095 & 3838 & 65.2 & 0.16 & 0.0277  & 0.91 & 833.3  & 421.6 & This work\\
Rb$_2$ 
  & 7.962     & 3933        & 57.5   & 0.01   & 0.0223 & 0 & 821.6 & 428.1 & This work\\
  & 7.9556(1) & 3993.593(3) & 57.75  & 0.125 & 0.0224 & 0 &       &       & \cite{AmiotJCP90,StraussPRA10} \\
RbCs 
  & 8.379  & 3750       & 50.1  & 0.11   & 0.0166 & 1.21      & 939.7 & 479.2 & This Work\\
  & 8.3660 & 3836.37(0) & 50.01 & 0.110 & 0.0166 & 1.23(1) &       &       & \cite{FellowsJMS99,DocenkoPRA11,MolonyPRL14} \\
RbFr 
  & 8.370 & 3619 & 46.0 & 0.14 & 0.0140 &  0.29 & 891.8 & 452.8 & This work\\
Cs$_2$ 
  & 8.801  & 3538      & 42.0  & 0.081 & 0.0117 & 0 & 1061.0 & 538.4 & This work\\
  & 8.7781 & 3649.9(5) & 42.021 & 0.082 & 0.0117 & 0 &        &       & \cite{AmiotJCP02}\\
CsFr 
  & 8.779 & 3440 &  37.7 & 0.077 & 0.0094 & 0.88 & 1008.5 & 507.8 & This work\\
Fr$_2$
  & 8.768 & 3333 & 32.8 & 0.04 & 0.0070 & 0     & 953.9 & 477.6 & This work\\
\end{tabular*}
\label{table:AM-AM-S}
\end{ruledtabular}
\end{table*}

\begin{table*}[tbp]
\begin{ruledtabular}
\caption{Spectroscopic constants of the alkali-metal diatomic molecules in the first triplet $a^3\Sigma^+$ electronic state: equilibrium interatomic distance $R_e$, well depth $D_e$, transition energies $T_e$, harmonic constant $\omega_e$, first anharmonicity constant $\omega_ex_e$, rotational constant $B_e$, permanent electric dipole moments $d_e$, and parallel $\alpha_e^\parallel$ and perpendicular $\alpha_e^\perp$ components of the static electric dipole polarizability. Available experimental values are also collected.} 
\begin{tabular*}{\textwidth}{lrrrrrrrrrrr}
Molecule & $R_e\,$(bohr)& $D_e\,$(cm$^{-1}$) & $T_e\,$(cm${}^{-1}$) & $\omega_e\,$(cm$^{-1}$) & $\omega_e\chi_e\,$(cm$^{-1}$)&  $B_e\,$(cm$^{-1}$) & $d_e\,$(D) & $\alpha_e^\parallel\,$(a.u.) & $\alpha_e^\perp\,$(a.u.) & Ref \\
 \hline
Li$_2$ 
  & 7.880      & 333.6        & 8175   & 65.1   & 3.31   & 0.2768 & 0 & 696.0 & 252.4 & This work\\
  & 7.88017(6) & 333.780(6) & 8183.0 & 65.2(2) & 3.19(4) & 0.2764 & 0 &       &       & \cite{MartinSAA88,SemczukPRA13} \\
LiNa 
  & 8.910 & 238      & 6847 &  42.12 & 1.80 & 0.1413 & 0.18 & 560.7 & 270.8 & This work\\
  &       & 229.753  & 6874 &  & & &  & & & \cite{RvachovPCCP18} \\
LiK 
  & 9.402    & 283    &  5914    & 43.5      & 1.70 & 0.1146 & 0.32 & 817.8 & 369.9 & This work\\
  & 9.43(17) & 287(4) &  5930(4) & 44.2(1.5) & & & & & & \cite{SalamiJCP07}\\
LiRb 
  & 9.718   & 276    & 5626     & 41.0 & 1.60 & 0.0984 & 0.37 & 845.2 & 399.1 & This work\\
  & 9.71(2) & 278(4) & 5650(8) &  & & & & & & \cite{IvanovaJCP11a}\\
LiCs  
  & 9.991 & 295     & 5522     & 40.7  & 1.30 &  0.0905 & 0.45 & 985.2 & 461.5 & This work\\
  & 9.916 & 309(10) & 5566(10) & 44.51 & 1.25 & & & & & \cite{StaanumPRA07}\\
LiFr 
  & 10.239 & 250 & 5209 & 37.7 & 1.5 & 0.0844 & 0.21 & 813.0 & 415.7 & This work\\
Na$_2$ 
  & 9.736    & 185       & 5808   & 25.2  & 0.88   & 0.0552 & 0 & 491.6 & 282.0 & This work\\
  & 9.719(2) & 172.91(4) & 5849.1 & 23.79 & 0.643 & 0.0565 & 0 &       &       & \cite{LiJCP85,KnoopPRA11}\\
NaK  
  & 10.301 & 206       & 5042   & 23.5  & 0.97  & 0.0392 & 0.03 & 697.9 & 387.3 & This work\\
  & 10.280 & 207.86(2) & 5065.8 & 23.01 & 0.622 & 0.0394 & & & & \cite{GerdesEPJD08}\\
NaRb 
  & 10.615 & 201       & 4792   & 19.9  & 0.47 & 0.0295 & 0.06 & 728.0 & 416.0 & This work\\
  & 10.583 & 203.36(5) & 4827.1 & 18.86 &      & 0.0282 &      &       &       & \cite{WangJCP91,PashovPRA05} \\
NaCs 
  & 10.902 & 212       & 4686   & 19.30 & 0.44  & 0.0259 & 0.09 & 837.71 & 482.0  & This work\\
  & 10.856 & 217.17(1) & 4737.1 & 19.57 & 0.44 & 0.0261 & & & & \cite{DocenkoJPB06}\\
NaFr 
  & 11.092 & 186 & 4413 & 17.75 & 0.40 & 0.0235 & 0.10 & 710.4 & 430.6 & This work\\
K$_2$ 
  & 10.853     & 248       & 4167   & 21.0    & 0.44 & 0.0262 & 0 & 946.3 & 483.8 & This work\\
  & 10.8364(2) & 255.02(5) & 4195.7 & 21.63(4) & 0.470 & 0.0260 & 0 & & & \cite{LiJCP90,PashovEPJD08}\\
KRb 
  & 11.211     & 242       & 3928   & 19.32  & 0.39 & 0.0180 & 0.06 & 972.9 & 512.5 & This work\\
  & 11.1549(2) & 249.03(1) & 3968.8 & 18.79 & 0.98 & 0.0181 &      &       &       & \cite{PashovPRA07}\\
KCs 
  & 11.475     & 256       & 3740   & 16.5  & 0.24 & 0.0152 & 0.10 & 1096.1 & 576.0 & This work\\
  & 11.4355(2) & 267.14(2) & 3803.0 & 17.52 & 0.306 & 0.0158 &      &        &       & \cite{FerberJCP08,FerberPRA09}\\
KFr
  & 11.685& 220& 3617 & 14.8 & 0.26 & 0.0133 &  0.13 & 935.5 & 530.4 & This work\\
Rb$_2$ 
  & 11.491     & 235        & 3698   & 13.4  & 0.16  & 0.0107 & 0 & 999.0 & 541.3 & This work\\
  & 11.5160(2) & 241.503(3) & 3752.1 & 13.50 & 0.171 & 0.0108 & 0 &  & & \cite{BeserJCP09,StraussPRA10}\\
RbCs 
  & 11.810     & 249       & 3501   & 12.29 & 0.15  & 0.0083  & 0.03 & 1118.2 & 604.9 & This work\\
  & 11.7528(2) & 259.34(3) & 3577.0 & 12.55 & 0.15 & 0.00841 & & & & \cite{DocenkoPRA11}\\
RbFr 
  & 12.010 & 215 & 3535  & 10.3 & 0.03 & 0.0068 & 0.07 & 961.3 & 558.8 & This work\\
Cs$_2$ 
  & 12.143    & 263       & 3275   & 10.9  & 0.082  & 0.0062  & 0 & 1237.4 & 668.2 & This work\\
  & 11.915(2) & 278.58(4) & 3370.7 & 11.37 & 0.184  & 0.00652 & 0 &        &       & \cite{XieJCP09,LauJCP16}\\ 
CsFr 
  & 12.344 & 226 & 3215 & 9.20 & 0.11 & 0.0048 & 0.07 & 1073.6 & 626.8 & This work\\
Fr$_2$ 
  & 12.502 & 198 & 3135  & 7.37 & 0.04 & 0.0035 & 0 & 929.3 & 576.0 & This work\\
\end{tabular*}
\label{table:AM-AM-T}
\end{ruledtabular}
\end{table*}

\begin{table*}[!tb]
\begin{ruledtabular}
\caption{Spectroscopic constants of the alkali-metal--alkaline-earth-metal diatomic molecules in the ground  $X^2\Sigma^+$ electronic state: equilibrium interatomic distance $R_e$, well depth $D_e$, harmonic constant $\omega_e$, first anharmonicity constant $\omega_ex_e$, rotational constant $B_e$, permanent electric dipole moments $d_e$, and parallel $\alpha_e^\parallel$ and perpendicular $\alpha_e^\perp$ components of the static electric dipole polarizability. Available experimental values are also collected.} 
\begin{tabular*}{\textwidth}{lrrrrrrrrr}
Molecule & $R_e\,$(bohr)& $D_e\,$(cm${}^{-1}$) & $\omega_e\,$(cm${}^{-1}$) & $\omega_e\chi_e\,$(cm${}^{-1}$)&  $B_e\,$(cm${}^{-1}$) & $d_e\,$(D) & $\alpha_e^\parallel\,$(a.u.) & $\alpha_e^\perp\,$(a.u.) & Ref \\
\hline
LiBe
   & 4.86 & 2545      & 311.8      & 5.2   & 0.6466 & 3.60 & 371.9  & 111.6 & This work\\
   &      & 2794(143) & 326(4) & 9.5(7) &  & & & & \cite{PersingerJPCA21}\\
NaBe 
   & 5.62 & 1293 & 166.2    & 3.6  & 0.2943 & 2.21 & 397.6 & 136.5 & This work\\
KBe 
   & 6.63& 898 & 120.3 & 4.9 & 0.1871 & 1.99 & 629.7 & 246.4 & This work\\
RbBe 
   & 7.02& 778 & 99.1 & 3.1 & 0.1501 & 1.74 & 657.2 & 277.5 & This work\\
CsBe  
   & 7.45 & 720 & 91.4 & 3.4 & 0.1287 & 1.72 & 784.3 & 336.5 & This work\\
FrBe  
   & 7.63 & 620 & 83.1 & 3.3 & 0.1194  & 1.27 & 621.3 & 297.3 & This work\\
LiMg 
   & 5.83  & 1561     & 185.1 & 6.1 & 0.3262 & 1.04 & 484.7 & 169.1 & This work\\
   & 5.86  & 1424(21) & 187.3  & 6.16 & 0.320 & & & & \cite{PersingerJPCA21b,BerryCPL97}\\
NaMg  
   & 6.61 & 910 &  87.3  & 2.5 & 0.1173 & 0.73 & 369.7 & 183.8 & This work\\
KMg 
   & 7.63& 722 & 61.5 & 1.61 & 0.0696 & 0.86 & 631.2 & 292.0 & This work\\
RbMg 
   & 8.01& 658 & 50.3  & 1.04 & 0.0502  & 0.83 & 659.2 & 321.0 & This work\\
CsMg  
   & 8.43 & 631 & 46.1 & 1.01 & 0.0417 & 0.88 & 772.1 & 383.5 & This work\\
FrMg  
   & 8.57 & 569 & 42.3  & 0.91 & 0.0379 & 0.70 & 635.9 & 335.4 & This work\\
LiCa 
   & 6.34      & 2553     & 203.1  & 4.3  & 0.2510  & 1.15 & 582.5 & 228.1 & This work\\
   & 6.3419(2) & 2605(10) & 202.239 & 3.405 & 0.25071 & & &  & \cite{IvanovaJCP11}\\
NaCa  
   & 6.97 & 1636 & 100.0 & 1.53 & 0.0849 & 1.15 & 582.2 & 240.2 & This work\\
KCa 
   & 8.03  & 1250 & 66.7  & 0.90   & 0.0474  & 1.90 & 879.3 & 328.0 & This work\\
   & 8.011 &      & 67.983 & 0.0940 & 0.04754 & & & & \cite{GerschmannPRA17}\\
RbCa  
   & 8.40 &  1131 &  52.8  & 0.69 & 0.0314 & 1.89 & 935.7 & 353.4 & This work\\
CsCa 
   & 8.84 &  1085 & 47.0 & 0.59 & 0.0251 & 2.18 & 1113.6 & 403.5 & This work\\
FrCa  
   & 8.95 & 965 & 42.1 & 0.46 & 0.0222 & 1.71 & 927.0 & 340.9 & This work\\
LiSr 
   & 6.69 &  2473 & 181.5 & 2.9  & 0.0207  & 0.39 & 620.3 & 269.7 & This work \\
   & 6.699 &  -   & 182.93 & 3.026 & 0.02072 & -    & -      & -      & \cite{SchwankeJPB17} \\
NaSr 
   & 7.32 & 1612 & 91.0 & 4.3 & 0.0617 & 0.57 & 630.8 & 281.5 & This work\\
KSr 
   & 8.40 & 1222 & 54.1     & 0.67    & 0.0316  & 1.40 & 927.6 & 368.4 & This work\\
   &      &      & 54.6(1.0) & 0.56(2) & 0.0322 & & & & \cite{SzczepkowskiJQSRT18}\\
RbSr 
   & 8.78  & 1108 & 39.4   & 0.35   & 0.0181 & 1.50 & 980.7 & 393.9 & This work\\
   & 8.683 & 1158(20) & 40.3(7) & 0.4(1) & 0.0185 & & & & \cite{CiameiPCCP18}\\
CsSr
   & 9.24 & 1060 & 33.6 & 0.22 & 0.0133 & 1.79 & 1151.2 & 445.3 & This work\\
FrSr 
   & 9.34 & 952 &  29.6 & 0.31 & 0.0110  & 1.41 & 971.2 & 411.0 & This work\\
LiBa  
   & 6.87   & 3335 & 183.3  & 2.6  & 0.1908   & 0.51 & 596.4 & 347.5 & This work\\
   & 6.8395 &      & 196.282 & 2.313 & 0.19277 & & & & \cite{StringatJMS94}\\
NaBa  
   & 7.39 & 2198 & 94.1 & 1.21 & 0.0560 & 0.08 & 637.3 & 363.0 & This work\\
KBa  
   & 8.56 & 1539 &  56.22 & 0.52 & 0.0270 & 1.58 & 1014.2 & 423.0 & This work\\
RbBa  
   & 8.96 & 1387 & 43.5 & 0.84 & 0.0143 & 1.76 & 1092.2 & 452.3 & This work\\
CsBa  
   & 9.45 & 1292 & 32.0 & 0.12 & 0.0100 & 2.19 & 1288.2 & 497.8 & This work\\
FrBa 
   & 9.52 & 1165 & 27.3 & 0.06 & 0.0078 & 1.71 & 1106.3 & 468.7 & This work\\
LiRa  
   & 7.23 &  2152 & 157.6 & 2.5 & 0.1692 &  0.89 & 686.6 & 328.2 & This work\\
NaRa  
   & 7.87 & 1408  & 70.4 & 0.98 & 0.0466 & 0.35 & 696.5 & 336.9 & This work\\
KRa 
   & 9.03 &  1053 & 42.40 & 0.41 & 0.0222 & 0.50 & 954.6 & 428.2 & This work\\
RbRa  
   & 9.42 & 957 & 28.9 & 0.24 & 0.0110 & 0.65 & 997.7 & 453.9 & This work\\
CsRa  
   & 9.90 & 910 & 23.4 & 0.15 & 0.0074 & 0.87 & 1141.2 & 510.2 & This work\\
FrRa 
   & 9.98 & 831 & 19.3 & 0.037 & 0.0054 & 0.68 & 980.4 & 470.2 & This work\\
\end{tabular*}
\label{table:AM-AEM}
\end{ruledtabular}
\end{table*}

\begin{table*}[tb]
\begin{ruledtabular}
\caption{Spectroscopic constants of the alkaline-earth-metal diatomic molecules in the ground $X^1\Sigma^+$ electronic state: equilibrium interatomic distance $R_e$, well depth $D_e$, harmonic constant $\omega_e$, first anharmonicity constant $\omega_ex_e$, rotational constant $B_e$, permanent electric dipole moments $d_e$, and parallel $\alpha_e^\parallel$ and perpendicular $\alpha_e^\perp$ components of the static electric dipole polarizability. Available experimental values are also collected.} 
\begin{tabular*}{\textwidth}{lrrrrrrrrr}
Molecule & $R_e\,$(bohr)& $D_e\,$(cm${}^{-1}$) & $\omega_e\,$(cm${}^{-1}$) & $\omega_e\chi_e\,$(cm${}^{-1}$)&  $B_e\,$(cm${}^{-1}$) & $d_e\,$(D) & $\alpha_e^\parallel\,$(a.u.) & $\alpha_e^\perp\,$(a.u.) & Ref \\ 
\hline
Be$_2$  
  & 4.62   & 919        & 270.3 & 25.2 & 0.6255 & 0 & 136.2 & 61.1 & This work\\
  & 4.620(1) & 934.8(3) & 270.7  & 26.0 & 0.623   & 0  &        &       & \cite{MeshkovJCP14,BondybeyCPL84} \\
BeMg 
  & 6.20 & 435 &  80.2  & 5.73 & 0.2390 & 0.17 & 173.3 & 94.9  & This work \\
BeCa  
  & 6.53 & 783 &  113.0 & 5.0 & 0.1918  & 0.39 & 320.7 & 169.2 & This work\\
BeSr  
  & 6.93 & 756 & 101.0 & 4.5 & 0.1535 & 0.44 & 374.4 & 208.6 & This work\\
BeBa  
  & 7.16 & 868 & 105.4 & 4.1 & 0.1387 & 0.60 & 489.8 & 279.0 & This work\\
BeRa 
  & 7.62 & 632 & 81.6  & 3.4 & 0.1198 & 0.46 & 427.7 & 259.2 & This work\\
Mg$_2$ 
  & 7.35  & 434     &  51.9    & 1.8    & 0.0930     &  0 & 214.6 & 124.9 & This work\\
  & 7.352 & 430.3(5) & 51.12(4) & 1.64(1) & 0.0929(1) &  0 & & & \cite{KnockelJCP13,BalfourCJP70}\\
MgCa  
  & 7.67 & 673 & 60.3   & 1.87     & 0.0683 & 0.07 & 375.5 & 195.7 & This work\\
  & 7.632 & 691.5(5) & 60.26(2)& 1.652(6) & 0.06896(2) & & & & \cite{AtmanspacherJCP85}\\
MgSr  
  & 8.01 & 676 & 50.2 & 0.57 & 0.0498 & 0.009 & 426.3 & 234.9 & This work\\
MgBa  
  & 8.25 & 758 & 51.4 & 0.87 & 0.0433 & 0.002 & 547.0 & 303.9 & This work\\
MgRa 
  & 8.63 & 614 & 43.8 & 0.88 & 0.0373 & 0.13  & 482.6 & 284.7 & This work\\
Ca$_2$ 
  & 8.13  & 1050       & 63.4    & 1.16    & 0.0456   & 0 & 574.1 & 260.0 & This work\\
  & 8.083 & 1102.08(9) & 64.93(1) & 1.07(3) & 0.0461(1) &  0 & & & \cite{AllardPRA02,BalfourCJP75}\\
CaSr 
  & 8.50 & 1046 & 50.8 & 0.24 & 0.0303 & 0.15 & 645.4 & 295.5 & This work\\
CaBa 
  & 8.76 & 1196 & 51.0 & 0.53 & 0.0253  & 0.21 & 804.1 & 359.6 & This work\\
CaRa 
  & 9.12 & 943 & 42.0 & 0.47 & 0.0213 & 0.43 & 711.1 & 343.1 & This work\\
Sr$_2$ 
  & 8.88      & 1046       & 39.6  & 0.43  & 0.0174  & 0 & 718.6 & 330.2 & This work\\
  & 8.8288(2) & 1081.64(2) & 40.328 & 0.399 & 0.01758 & 0 & & & \cite{SteinPRA08,SteinEPJD10}\\
SrBa  
  & 9.15 & 1191 & 37.4 & 0.34 & 0.0134 & 0.007 & 886.1 & 392.9 & This work\\
SrRa  
  & 9.50 & 945 & 28.4 & 0.20 & 0.0106 & 0.31 & 783.5 & 377.3 & This work\\
Ba$_2$ 
  & 9.43 & 1366 & 34.3   & 0.20 & 0.0098 & 0 & 1080.4 & 453.5 & This work\\
  &      &      & 33.2(2) & 0.5(2) &        & 0 & & &\cite{LebeaultJMS98}\\
BaRa 
  & 9.78 & 1068 & 26.7 & 0.21 & 0.0074 & 0.43 & 952.3 & 439.2 & This work\\
Ra$_2$ 
  & 10.13 & 858 &  20.5 & 0.14 & 0.0052 & 0 & 842.3 & 424.6 & This work\\
\end{tabular*}
\label{table:AEM-AEM}
\end{ruledtabular}
\end{table*}

For the $X^1\Sigma^+$ state, the RMSE of the well depths is 54$\,$cm$^{-1}$ (1.3$\,$\%). The smallest error of 8$\,$cm$^{-1}$ (0.1$\,$\%) is for Li$_2$ and the largest error of 112$\,$cm$^{-1}$ (3$\,$\%) is for Cs$_2$. Only for the heaviest Cs$_2$ dimer, the error is larger than 100$\,$cm$^{-1}$, and only for RbCs and Cs$_2$, the error is larger than 2$\,$\%. For most alkali-metal molecules in the $X^1\Sigma^+$ state, the error is smaller than 1$\,$\%. The RMSE of the equilibrium distances is 0.012$\,$bohr (0.16$\,$\%). The smallest error of 0.001$\,$bohr (0.03$\,$\%) is for Li$_2$ and the largest error of 0.02$\,$bohr (0.25$\,$\%) is for NaRb and Rb$_2$. The theoretical and experimental harmonic and anharmonicity constants agree with each other very well, mostly within 0.1-0.2$\,$cm$^{-1}$.

For the $a^3\Sigma^+$ state, the RMSE of the well depths is 8.5$\,$cm$^{-1}$ (3.5$\,$\%). The smallest error of $<1\,$cm$^{-1}$ ($<0.3\,$\%) is for Li$_2$ and the largest error of 16$\,$cm$^{-1}$ (5.7$\,$\%) is for Cs$_2$. The RMSE of the equilibrium distances is 0.07$\,$bohr (0.6$\,$\%). The smallest error of $<0.001\,$bohr ($<0.01\,$\%) is for Li$_2$ and the largest error of 0.22$\,$bohr (1.9$\,$\%) is for Cs$_2$. Molecules containing Cs have the largest errors, but the error is larger than 0.1$\,$bohr for the heaviest Cs$_2$ dimer only. The agreement for other spectroscopic constants is also good.

All the alkali-metal diatomic molecules have already been studied theoretically (see Table~\ref{tab:Ref_AM-AM}). The results for the $X^1\Sigma^+$ states, reported within the last two decades, typically agree with the experimental values within 200$\,$cm$^{-1}$ (5$\,$\%), with some of them having errors smaller than 100$\,$cm$^{-1}$ (3$\,$\%). For example, the calculations for 10 heteronuclear molecules in the $X^1\Sigma^+$ state reported in Ref.~\cite{FedorovJCP14} reached the RMSE of the well depths of 84$\,$cm$^{-1}$ and the RMSE of the equilibrium distances of 0.014$\,$bohr, which are by 55$\,$\% and 15 $\,$\% larger, respectively, than the RMSEs reported in this study. The results for the $a^3\Sigma^+$ states, reported within the last two decades, typically agree with the experimental values within 50$\,$cm$^{-1}$ (20$\,$\%), with some of them having errors smaller than 10$\,$cm$^{-1}$ (5$\,$\%). For example, the highly-accurate calculations for the lightest Li$_2$~\cite{LesiukPRA20} and NaLi~\cite{GronowskiPRA20} molecules in the $a^3\Sigma^+$ state reached recently the spectroscopic accuracy ($<1$cm$^{-1}$). The observed overall agreement of our results with the experimental values for the alkali-metal molecules confirms the accuracy and sufficiency of the employed methodology for describing the electronic structure of such molecules, is better on average than in previous theoretical studies but cross-validates the accuracy of different approaches, and allows to establish the present results as a benchmark and reference for future more accurate calculations.

\subsubsection{Alkali-metal--alkaline-earth-metal diatomic molecules}

The alkali-metal--alkaline-earth-metal diatomic molecules in their ground $X^2\Sigma^+$ electronic state are weakly to moderately bound and have the van der Waals character. Formally, they are chemically bound with a bond order of half. They are radicals because of their single unpaired valence electron. The calculated well depths range from 569$\,$cm$^{-1}$ for FrMg to 3335$\,$cm$^{-1}$ for LiBa (with an average of 1311$\,$cm$^{-1}$). They systematically decrease with increasing the atomic number of the alkali-metal atoms for a given alkaline-earth-metal atom, as clearly visible in Fig.~\ref{fig:AM-AEM}, but the trend is less regular with changing alkaline-earth-metal atoms. The molecules containing Li are significantly stronger bound than others. The corresponding equilibrium distances systematically increase with increasing the atomic numbers of both the alkali-metal and alkaline-earth-metal atoms and take moderate values between 4.86$\,$bohr for LiBe and 9.98$\,$bohr for RaFr (with an average of 7.94$\,$bohr). The alkali-metal--alkaline-earth-metal molecules have two-to-four times smaller (larger) well depth than the alkali-metal molecules in the $X^1\Sigma^+$ ($a^3\Sigma^+$) electronic state, but their equilibrium distances are only slightly larger than those of the ground-state alkali-metal dimers. 

Spectroscopic measurements of the rovibrational levels of the ground electronic state were recorded for the LiBe, LiMg, LiCa, KCa, LiSr, KSr, RbSr, and LiBa molecules (see Table~\ref{tab:Ref_AM-AEM}). Our spectroscopic constants are compared with the available experimental values in Table~\ref{table:AM-AEM}. However, the experimental data did not allow for accurate determination of well depths of corresponding potential energy curves. Nevertheless, our well depths agree within 10$\,$\% with the rough experimental estimates for LiBa~\cite{StringatJMS94} and LiMg~\cite{PersingerJPCA21b} obtained from the extrapolations of the Morse vibrational constants, and within 2$\,$\% and 5$\,$\% with the experimental estimates guided by previous theoretical calculations for LiCa~\cite{IvanovaJCP11} and RbSr~\cite{CiameiPCCP18}, respectively. The theoretical and experimental vibrational and rotational constants agree even better, mostly with differences below 3$\,$\%. Therefore, we estimate the uncertainty of the calculated potential energy curves for the alkali-metal--alkaline-earth-metal diatomic molecules to be around 3-6$\,$\%.

Most of the alkali-metal--alkaline-earth-metal diatomic molecules have already been studied theoretically (see Table~\ref{tab:Ref_AM-AEM}). Our results agree well with previous accurate calculations. For example, for the family of the LiSr, NaSr, KSr, RbSr, and CsSr molecules studied in Ref.~\cite{GueroutPRA10}, the average absolute differences between the present and previous well depths is 49$\,$cm$^{-1}$ (3.1$\,$\%) and equilibrium distances -- 0.11$\,$bohr (1.4$\,$\%), with all equilibrium distances systematically shorter in Ref.~\cite{GueroutPRA10}. Slightly worse agreement with the average absolute differences of 90$\,$cm$^{-1}$ (6.7$\,$\%) for well depths and 0.07$\,$bohr (0.9$\,$\%) for equilibrium distances is found when our results are compared with another systematic study of 16 alkali-metal--alkaline-earth-metal combinations presented in Ref.~\cite{PototschnigPCCP16}. It is worth mentioning that different methods, basis sets, and pseudopotentials were used in the present work and Refs.~\cite{GueroutPRA10,PototschnigPCCP16}, thus the observed overall agreement additionally cross-validates the accuracy of different approaches. The differences are a bit larger between the present and some older results, which used smaller basis sets and less accurate wave functions.

\begin{figure*}
\begin{center}
\includegraphics[width=\textwidth]{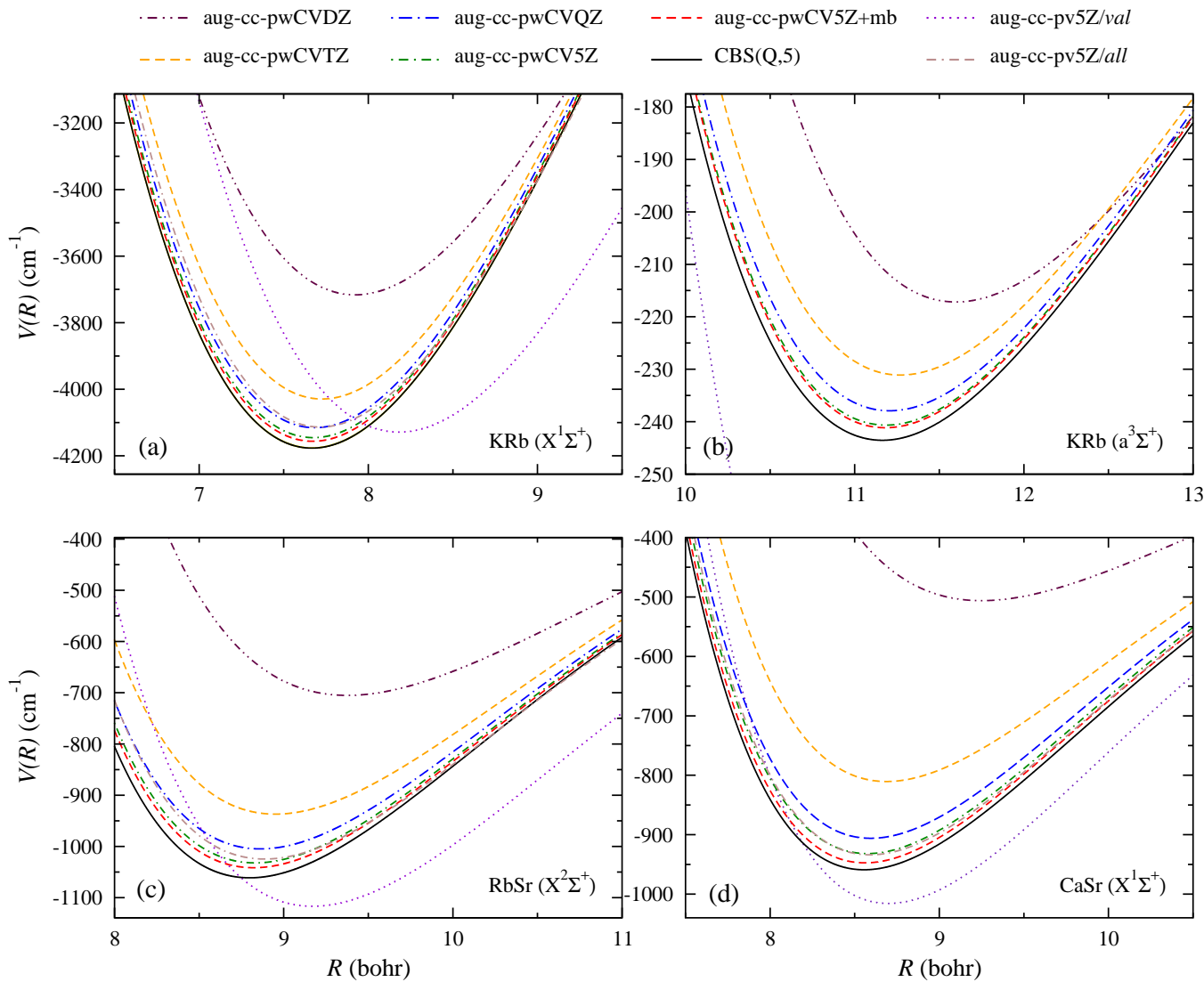}
\caption{Potential energy curves of (a) the KRb molecule in the $X^1\Sigma^+$ electronic state, (b) the KRb molecule in the $a^3\Sigma^+$ electronic state, (c) the RbSr molecule in the $X^2\Sigma^+$ electronic state, and (d) the CaSr molecule in the $X^1\Sigma^+$ electronic state computed using different basis sets and the CCSD(T) method. See the text for details.}
\label{fig:basis-sets}
\end{center}
\end{figure*}

\subsubsection{Alkaline-earth-metal diatomic molecules}

The alkaline-earth-metal diatomic molecules in their ground $X^1\Sigma^+$ electronic state are weakly bound van der Waals complexes. Formally, they are not chemically bound with a bond order of zero and are stabilized by the dispersion interaction. The calculated well depths range from 434$\,$cm$^{-1}$ for Mg$_2$ to 1366$\,$cm$^{-1}$ for Ba$_2$ (with an average of 869$\,$cm$^{-1}$). The trends are not regular with changing the atomic number of the alkaline-earth-metal atoms (see Fig.~\ref{fig:AEM-AEM}), but generally heavier atoms are stronger bound due to their larger polarizability, as expected for van der Waals systems, with an exception for the unusually strongly bound Be$_2$ dimer~\cite{MerrittScience09}. The corresponding equilibrium distances systematically increase with increasing the atomic number of the alkaline-earth-metal atoms and take moderate values between 4.62$\,$bohr for Be$_2$ and 10.13$\,$bohr for Ra$_2$ (with an average of 8.11$\,$bohr). Thus, the alkaline-earth-metal molecules have slightly smaller well depths and slightly larger equilibrium distances than the corresponding alkali-metal--alkaline-earth-metal molecules.   

Spectroscopic measurements of the rovibrational levels of the ground electronic state were recorded for the Be$_2$, Mg$_2$, MgCa, Ca$_2$, Sr$_2$, and Ba$_2$ molecules (see Table~\ref{tab:Ref_AEM-AEM}), and the accurate dissociation energies were determined for the listed molecules except for Ba$_2$. Our spectroscopic constants are compared with the available experimental values in Table~\ref{table:AEM-AEM}. The RMSE of the well depths is 30$\,$cm$^{-1}$ (3.0\%). The smallest error of 4$\,$cm$^{-1}$ (0.9$\,$\%) is for Mg$_2$ and the largest error of 52$\,$cm$^{-1}$ (4.7$\,$\%) is for Ca$_2$. The RMSE of the equilibrium distances is 0.03$\,$bohr (0.4$\,$\%). The theoretical and experimental vibrational and rotational constants also mostly agree within 3\%. Therefore, we estimate the uncertainty of the calculated potential energy curves for the alkaline-earth-metal diatomic molecules to be around 3-6\% with possibly larger values for heavier molecules. 

Most of the alkaline-earth-metal diatomic molecules have already been studied theoretically (see Table~\ref{tab:Ref_AEM-AEM}), but the detailed or accurate calculations have been reported for homonuclear Be$_2$, Mg$_2$, Ca$_2$, Sr$_2$, and Ba$_2$ dimers, only, while heteronuclear combinations were studied using small basis sets and low-level methods. The highly-accurate calculations reaching the spectroscopic accuracy ($<1$cm$^{-1}$) were presented for the lightest Be$_2$~\cite{PatkowskiScience09,LesiukJCTC19}, and recently Mg$_2$~\cite{YuwonoSA20}. Our results agree within 1$\,$\% with these more accurate approaches and establish the most accurate reference for all other alkaline-earth-metal molecules.

\subsection{Convergence and accuracy}

\begin{figure*}
\begin{center}
\includegraphics[width=\textwidth]{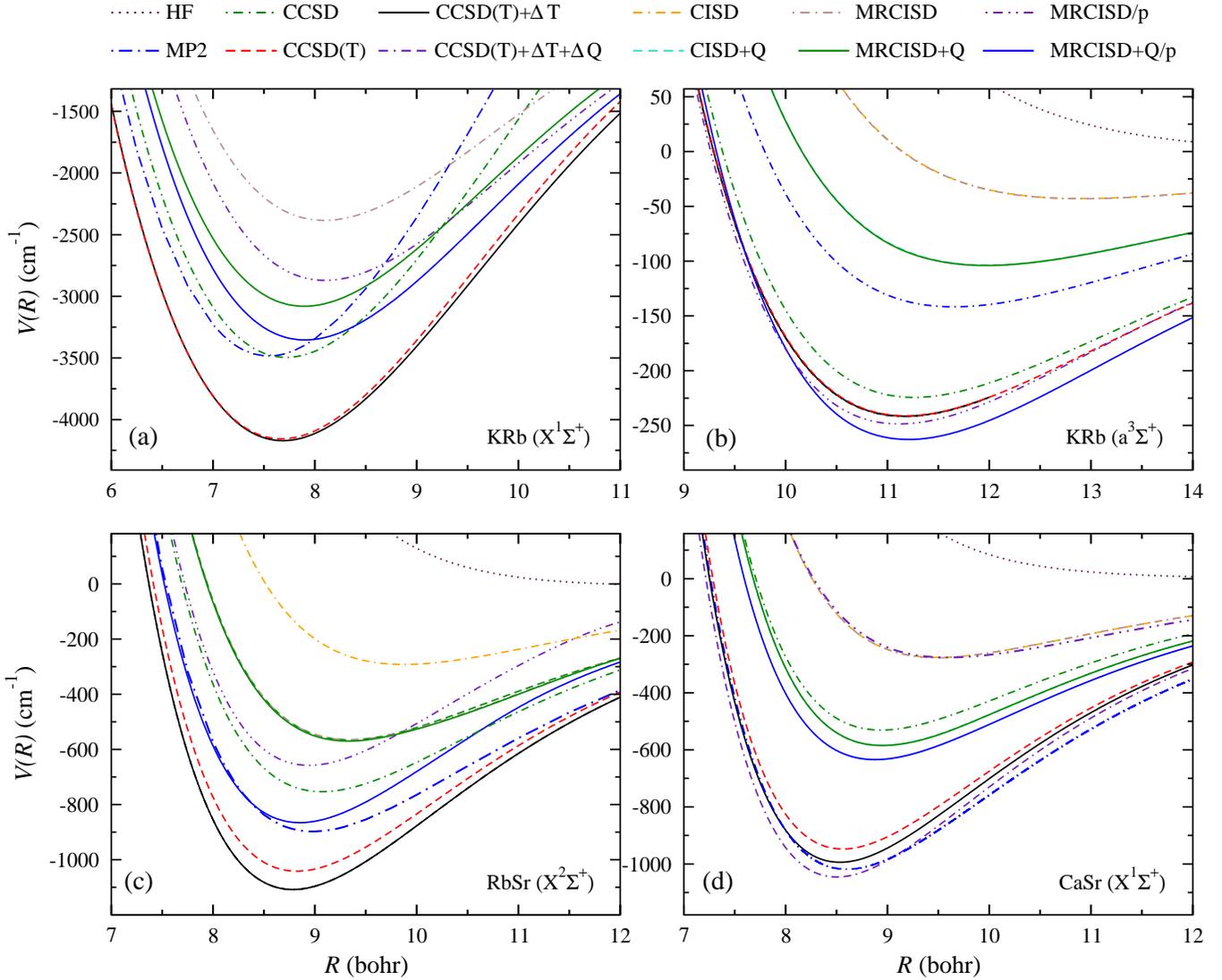}
\caption{Potential energy curves of (a) the KRb molecule in the $X^1\Sigma^+$ electronic state, (b) the KRb molecule in the $a^3\Sigma^+$ electronic state, (c) the RbSr molecule in the $X^2\Sigma^+$ electronic state, and (d) the CaSr molecule in the $X^1\Sigma^+$ electronic state computed with the different electronic-structure methods and the aug-cc-pCV5Z basis sets. See the text for details.}
\label{fig:methods}
\end{center}
\end{figure*}

Estimating the theoretical uncertainties of molecular electronic structure calculations is challenging. A prerequisite is an accurate description of the involved atoms. The performance of the CCSD(T) method with the aug-cc-pwCV5Z basis sets for obtaining properties of the alkali-metal and alkaline-earth-metal atoms was carefully tested in Refs.~\cite{HillJCP17,SmialkowskiPRA21} and demonstrated to reproduce well the most accurate available theoretical or experimental data. For example, the calculated atomic static electric dipole polarizabilities coincide with previous accurate values with the RMSE of 3$\,e^2a_0^2/E_\text{h}$~(1\%), and the atomic ionization potentials and the lowest $S$--$P$ excitation energies agree with experimental data with the RMSE of 172$\,$cm$^{-1}$ (0.5$\,$\%) and 109$\,$cm$^{-1}$ (0.8$\,$\%), respectively.

In this subsection, we analyze the convergence of the interatomic interaction energy calculations with the size of the employed atomic basis sets and the quality of the employed wave-function representation for the selected exemplary KRb, RbSr, and CaSr molecules.

Figure~\ref{fig:basis-sets} presents the potential energy curves of the KRb molecule in the $X^1\Sigma^+$ and $a^3\Sigma^+$ electronic states, the RbSr molecule in the $X^2\Sigma^+$ electronic state, and the CaSr molecule in the  $X^1\Sigma^+$ electronic state calculated with several different basis sets and the CCSD(T) method. The family of the aug-cc-pwCV$n$Z basis sets is employed, including the largest aug-cc-pwCV5Z basis sets augmented by bond functions (aug-cc-pwCV5Z+bf). We also show the interaction energies obtained with the two largest basis sets, with $n$=Q and 5 extrapolated to the complete basis set limit (CBS) using the $1/n^3$ two-point formula~\cite{HelgakerJCP97}. Additionally, the aug-cc-pV5Z basis sets with all electrons correlated (aug-cc-pV5Z/all) and with only valence electrons correlated (aug-cc-pV5Z/val) are used.  

The smooth convergence toward the complete basis set limit can be observed, with the aug-pwCVTZ basis sets performing reasonably well and the aug-pwCVQZ basis sets being less than 5$\,$\% from the CBS limit. To reach an accuracy better than 5$\,$\%, the largest basis sets have to be used. The CBS limit can be approached using either the extrapolation scheme or bond functions. In our test calculations, the extrapolated potential energy curves have slightly larger (by 0.5-2$\,$\%) well depths than the curves obtained with the basis sets augmented by bond functions, with bond functions performing best for the alkaline-earth-metal dimer. Additionally, we studied the convergence of the CCSDT correction of Eq.~\eqref{eq:VSDT} with the basis set size and found the opposite trend.  In the final calculations, we decided to use the bond functions to accelerate the convergence toward the CBS limit because this approach is simpler and works very well for weakly-bound van der Waals complexes~\cite{TaoIRPC01}, including other metal-atom molecules~\cite{ZarembaPRA21}. 

The importance of including the inner-shell electron correlation is evident when the potential energy curves obtained using the aug-cc-pV5Z basis sets with and without correlating $(n-1)s^2(n-1)p^6$ electrons are compared. The equilibrium distances are larger by around 0.5$\,$bohr for all the studied molecules, and the well depths are significantly larger for all the molecules except ground-state alkali-metal dimers when only valence electrons are correlated. Thus, including the inner-shell electron correlation (core-core and core-valence contributions) is crucial for accurately describing the interatomic interactions in the studied systems. This correlation can be included directly, as in the present work, or effectively using core polarization potentials in calculations with large-core pseudopotentials, which provides results in good agreement with the present values~\cite{AymarJCP05}.

Figure~\ref{fig:methods} presents the potential energy curves of the KRb molecule in the $X^1\Sigma^+$ and $a^3\Sigma^+$ electronic states, the RbSr molecule in the $X^2\Sigma^+$ electronic state, and the CaSr molecule in the  $X^1\Sigma^+$ electronic state calculated at several different levels of theory: RHF, MP2, CISD, CISD+Q, MRCISD, MRCISD+Q, MRCISD+$p$, MRCISD+Q+$p$, CCSD, CCSD(T), CCSD(T)+$\Delta$T, and CCSD(T)+$\Delta$T+$\Delta$Q, as introduced in Sec.~\ref{sec:methods}. The aug-cc-pwCV5Z+bf basis sets are used in all calculations except for the CCSD(T)+$\Delta$T and CCSD(T)+$\Delta$T+$\Delta$Q curves, which are obtained with Eq.~\eqref{eq:Vint}, including corrections given by Eqs.~\eqref{eq:VSDT} and~\eqref{eq:VSDTQ}.  

The mean-field spin-restricted Hartree-Fock method (RHF) does not properly describe the studied molecules because the correlation energy is crucial for binding in all the systems. Low-level methods such as the second-order perturbation theory (MP2) or the configuration interaction method, including single and double excitations (CISD) are presented for completeness and poorly reproduce the correlation energy, although the curve at the MP2 level is accidentally close to the final result for the alkaline-earth-metal dimer. The multireference version of the configuration interaction method with the active space of valence orbitals ($ns+ns$ in MRCISD) describes better the ground-state alkali-metal and alkali-metal--alkaline-earth-metal molecules but does not improve the CISD results for other molecules, where only single spin configuration can be constructed from the valence orbitals. Increasing the size of the active space in the configuration interaction approach by including the lowest unoccupied $p$ orbitals ($nsnp+nsnp$ in MRCISD/$p$) noticeably improves its accuracy. The inclusion of the Davidson correction to the CISD and MRCISD results (CISD+Q and MRCISD+Q, respectively) brings them closer to the most accurate coupled cluster values. However, the MRCI approach is size-inconsistent, and its convergence toward the full configuration interaction results by increasing the CI active space is hard to control and limited by available computing power.

\begin{figure}
\begin{center}
\includegraphics[width=\columnwidth]{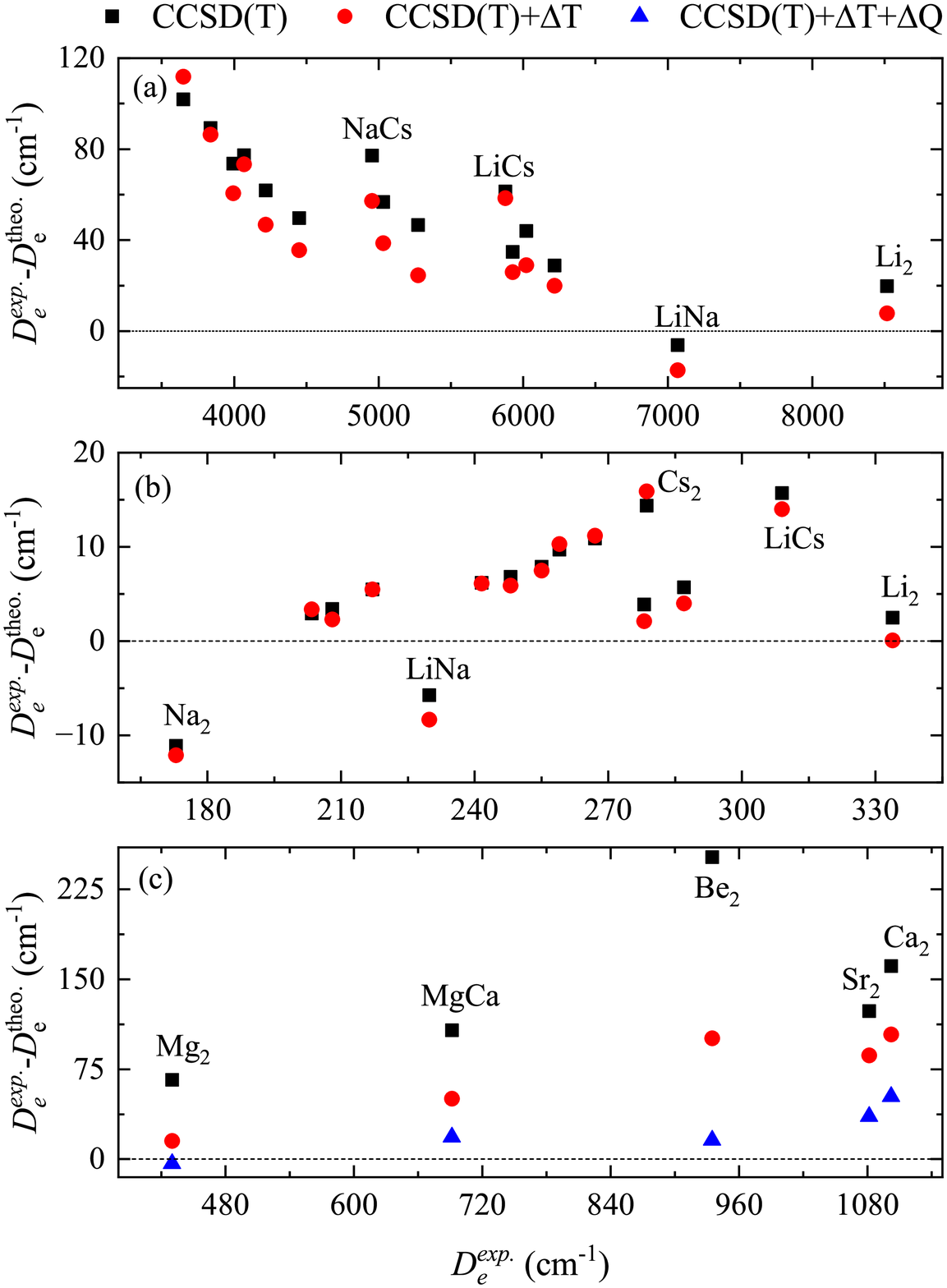}
\caption{Differences between the known accurate potential well depths $D_e^\text{exp.}$ and their present theoretical values $D_e^\text{theo.}$ obtained with the CCSD(T) method and including the CCSDT and CCSDTQ corrections as a function of the well depth for (a) the alkali-metal molecules in the ground $X^1\Sigma^+$ electronic state, (b) the alkali-metal molecules in the $a^3\Sigma^+$ electronic state, and (c) the alkaline-earth-metal molecules in the $X^1\Sigma^+$ electronic state. Labels are shown for selected molecules.}
\label{fig:CC_vs_exp}
\end{center}
\end{figure}

The coupled cluster (CC) method provides the most accurate interaction energies with a well-controlled convergence toward the full configuration interaction results by including higher and higher excitations in the CC wave function in a systematic way. It is also size-consistent. The inclusion of triple excitations is important for all the studied molecules. The so-called gold standard of quantum chemistry, the CCSD(T) method, which provides a good estimate of connected triple excitations perturbatively, performs very well for the alkali-metal molecules. For these molecules, the inclusion of full triple excitations beyond the CCSD(T) results (CCSD(T)+$\Delta$T) changes the potential well depth by less than $0.5\,\%$ [see also Fig.~\ref{fig:CC_vs_exp}(a,b)], and is more important at larger distances. In contrast, the CCSDT corrections increase the well depths for the alkali-metal--alkaline-earth-metal and alkaline-earth-metal molecules by more than $5\,\%$. The importance of full triple excitations for the molecules containing alkaline-earth-metal atoms is not surprising since the CCSDT and CCSDTQ methods are needed to describe the alkali-metal--alkaline-earth-metal and alkaline-earth-metal molecules at the valence full configuration level, respectively. Therefore, the CCSDTQ corrections increase the well depths for the alkaline-earth-metal molecules by an additional $5\,\%$ and is important to achieve accurate description [see also Fig.~\ref{fig:CC_vs_exp}(c)].

We assess the importance of higher excitations in the coupled cluster calculations based on statistics for all the studied molecules. The inclusion of the CCSDT corrections of Eq.~\eqref{eq:VSDT} increases on average the well depths for the alkali-metal molecules in the $X^1\Sigma^+$ state by 10$\,$cm$^{-1}$ (0.3$\,$\%) (except Cs$_2$, CsFr, Fr$_2$ with the opposite effect of the same magnitude), for the alkali-metal--alkaline-earth-metal molecules in the $X^2\Sigma^+$ state by 88$\,$cm$^{-1}$ (7.5$\,$\%), and for the alkaline-earth-metal molecules in the $X^1\Sigma^+$ state by 52$\,$cm$^{-1}$ (8.1$\,$\%). For the alkali-metal molecules in the $a^3\Sigma^+$ state, the well depths change on average by 1.1$\,$cm$^{-1}$ (0.4$\,$\%) with an increase (decrease) of binding energy for lighter (heavier) molecules. The inclusion of the CCSDT corrections also increases, on average, the equilibrium distances for the alkali-metal molecules in the $X^1\Sigma^+$ state by 0.017$\,$bohr (0.2$\,$\%) and decreases on average the equilibrium distances for the alkali-metal molecules in the $a^3\Sigma^+$ state by $<0.01\,$bohr ($<0.1\,$\%), for the alkali-metal--alkaline-earth-metal molecules in the $X^2\Sigma^+$ by 0.04$\,$bohr (0.5$\,$\%), and for the alkaline-earth-metal molecules in the $X^1\Sigma^+$ state by 0.1$\,$bohr (2.0$\,$\%). Additionally, the CCSDTQ corrections of Eq.~\eqref{eq:VSDTQ} increase on average the well depths for the alkaline-earth-metal molecules in the $X^1\Sigma^+$ state by 44$\,$cm$^{-1}$ (5.1$\,$\%) and decrease on average their equilibrium distances by 0.03$\,$bohr (0.4$\,$\%). The importance of higher excitations in the coupled cluster calculations in reproducing the known experimental potential well depths is presented in Fig.~\ref{fig:CC_vs_exp}, which confirms that the description of valence electrons at the full configuration interaction level is essential to achieve accurate predictions.

\begin{figure*}[t]
\begin{center}
\includegraphics[width=\textwidth]{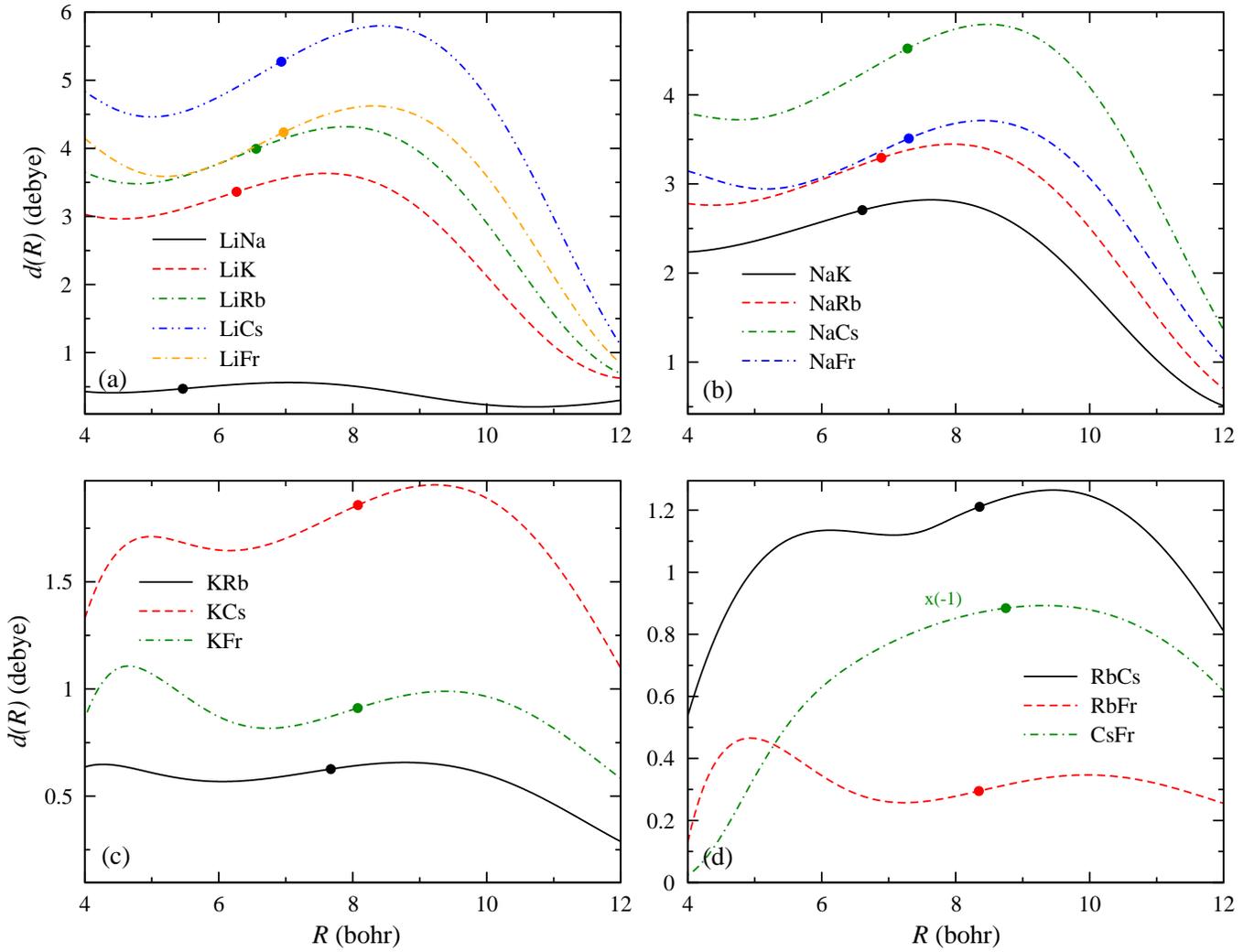}
\caption{Permanent electric dipole moment curves of the alkali-metal diatomic molecules in the $X^1\Sigma^+$ electronic state. The points indicate values for equilibrium distances.}
\label{fig:dm-AM-AM-S}
\end{center}
\end{figure*}

\begin{figure*}[t]
\begin{center}
\includegraphics[width=\textwidth]{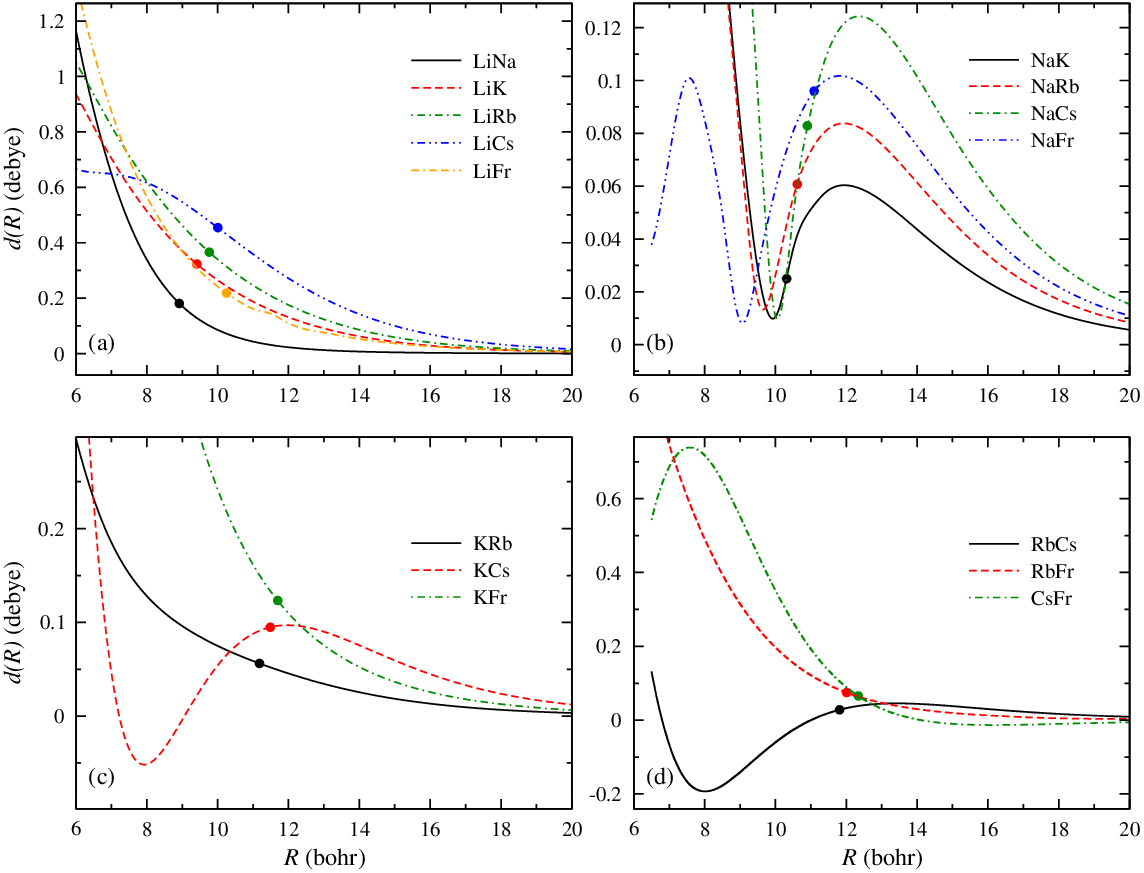}
\caption{Permanent electric dipole moment curves of the alkali-metal diatomic molecules in the $a^3\Sigma^+$ electronic state. The points indicate values for equilibrium distances.}
\label{fig:dm-AM-AM-T}
\end{center}
\end{figure*}

Based on the convergence analysis in this subsection and the comparison with available experimental data in the previous subsection, where a good agreement was observed, we can conclude that the employed electronic structure methods, basis sets, and energy-consistent pseudopotentials properly treat the relativistic effects and reproduce the correlation energy while being close to being converged in the size of the basis function set. We can also estimate the uncertainties of the employed computational approach of Eq.~\eqref{eq:Vint} for calculating potential energy curves are around:
\begin{itemize}  
\item 25-100$\,$cm$^{-1}$ (0.5-2$\,$\%) of $D_e$ and 0.005-0.02$\,$bohr (0.05-0.2$\,$\%) of $R_e$ for the alkali-metal molecules in the $X^1\Sigma^+$ state,
\item 5-15$\,$cm$^{-1}$ (2-6$\,$\%) of $D_e$ and 0.05-0.2$\,$bohr (0.5-2$\,$\%) of $R_e$ for the alkali-metal molecules in the $a^3\Sigma^+$ state,
\item 50-100$\,$cm$^{-1}$ (3-6$\,$\%) of $D_e$ and 0.01-0.04$\,$bohr (0.1-0.5$\,$\%) of $R_e$ for the alkali-metal--alkaline-earth-metal molecules in the $X^2\Sigma^+$ state,
\item 50-100$\,$cm$^{-1}$ (3-6$\,$\%) of $D_e$ and 0.01-0.05$\,$bohr (0.1-0.6$\,$\%) of $R_e$ for the alkaline-earth-metal molecules in the $X^1\Sigma^+$ state.
\end{itemize}

\begin{figure*}[t]
\begin{center}
\includegraphics[width=\textwidth]{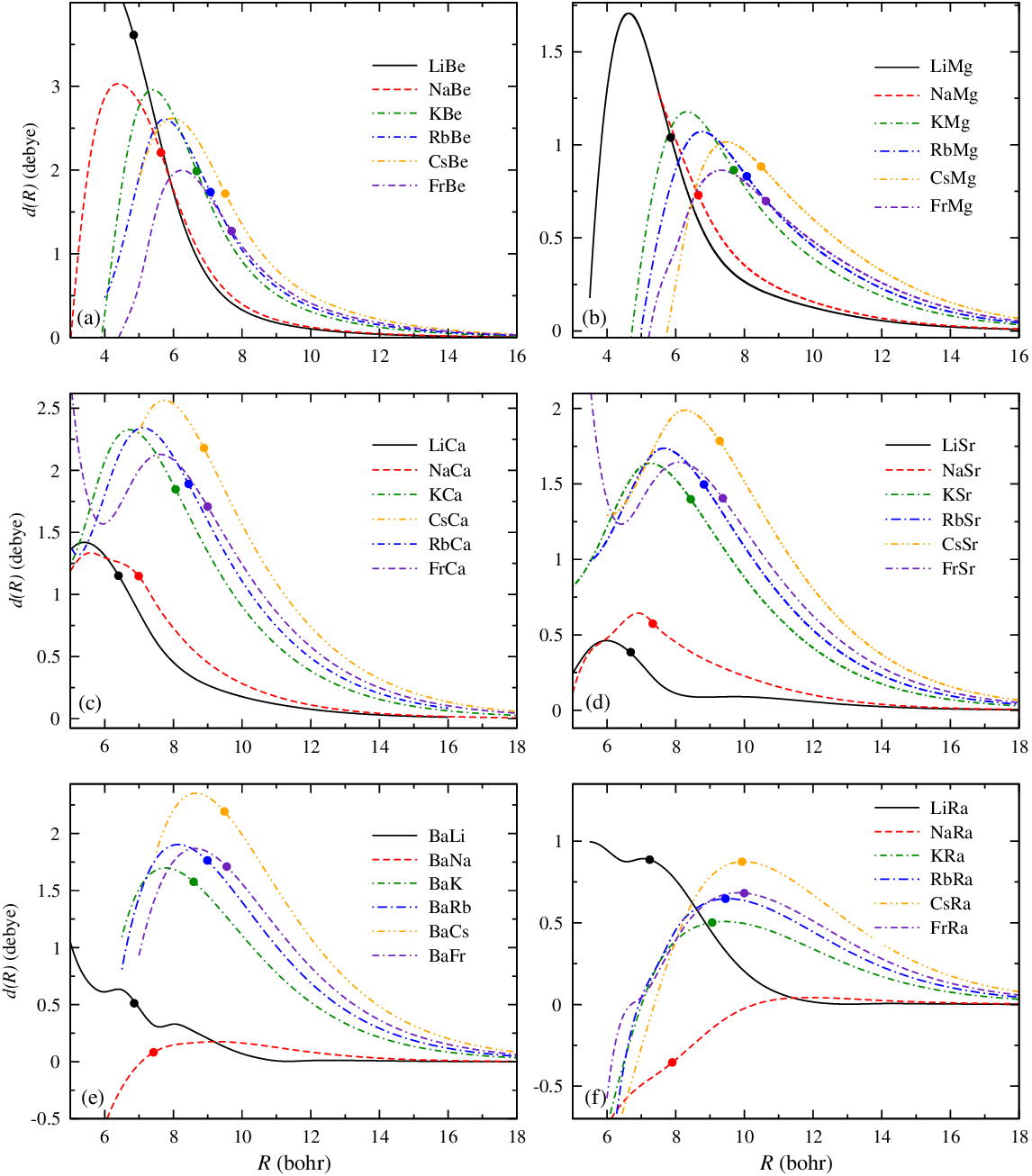}
\caption{Permanent electric dipole moment curves of the alkali-metal--alkaline-earth-metal diatomic molecules in the $X^2\Sigma^+$ electronic state. The points indicate values for equilibrium distances.}
\label{fig:dm-AM-AEM}
\end{center}
\end{figure*}
  
\begin{figure*}[t]
\begin{center}
\includegraphics[width=\textwidth]{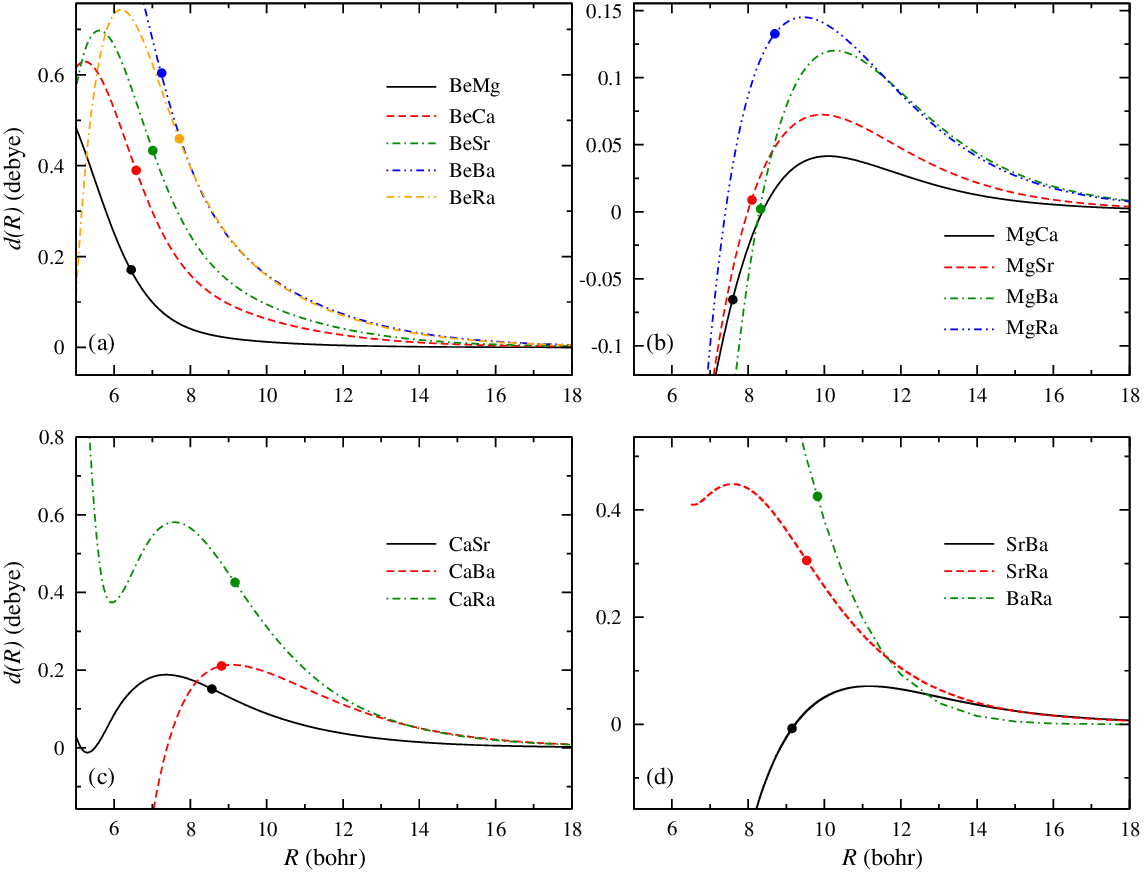}
\end{center}
\caption{Permanent electric dipole moment curves of the alkaline-earth-metal diatomic molecules in the $X^1\Sigma^+$ electronic state. The points indicate values for equilibrium distances.}
\label{fig:dm-AEM-AEM}
\end{figure*}  

\subsection{Permanent electric dipole moments}

The computed permanent electric dipole moments as functions of the internuclear distance for the alkali-metal diatomic molecules in the $X^1\Sigma^+$ and $a^3\Sigma^+$ electronic states are presented in Fig.~\ref{fig:dm-AM-AM-S} and Fig.~\ref{fig:dm-AM-AM-T}, respectively, for the alkali-metal--alkaline-earth-metal diatomic molecules in the $X^2\Sigma^+$ electronic state -- in Fig.~\ref{fig:dm-AM-AEM}, and for the alkaline-earth-metal diatomic molecules in the $X^1\Sigma^+$ state -- in Fig.~\ref{fig:dm-AEM-AEM}. Calculations are performed for all the combinations of the alkali-metal (Li, Na, K, Rb, Cs, Fr) and alkaline-earth-metal (Be, Mg, Ca, Sr, Ba, Ra) atoms. All curves exhibit smooth behavior, and different classes of molecules share similar characteristics. The corresponding values for the equilibrium distances, $d_e\equiv d(R_e)$, are indicated in the plots and collected in Tables~\ref{table:AM-AM-S}-\ref{table:AEM-AEM}, along with available experimental data.
  
The permanent electric dipole moments originate from the uneven distribution of charge in heteronuclear molecules. They describe the response of polar molecules to external static electric fields, resulting in molecular polarization~\cite{LemeshkoMP13}. The dipole-dipole interactions also dominate long-range intermolecular interactions and ultracold collisions between polar molecules and are crucial for their applications~\cite{GadwayJPB16}.

The alkali-metal diatomic molecules in their ground $X^1\Sigma^+$ electronic state possess the largest permanent electric dipole moments ranging from 0.3$\,$D for RbFr to 5.3$\,$D for LiCs (with an average of 2.5$\,$D) at their equilibrium distances. Molecules containing Li are the most polar, followed by those containing Na. In contrast, the alkali-metal diatomic molecules in the $a^3\Sigma^+$ electronic state have the smallest permanent electric dipole moments ranging from 0.03$\,$D for NaK and RbCs to 0.45$\,$D for LiCs (with an average of 0.15$\,$D) at their equilibrium distances. Triplet-state molecules containing Li are also the most polar.

The alkali-metal--alkaline-earth-metal diatomic molecules in their ground $X^2\Sigma^+$ electronic state exhibit intermediate permanent electric dipole moments ranging from 0.08$\,$D for NaBa to 3.6$\,$D for LiBe (with an average of 1.3$\,$D) at their equilibrium distances. Among them, molecules containing Be or Cs are the most polar. Finally, the alkaline-earth-metal diatomic molecules in their ground $X^2\Sigma^+$ electronic state have small permanent electric dipole moments ranging from almost zero for combinations containing Mg to 0.6$\,$D for BeBa (with an average of 0.25$\,$D) at their equilibrium distances. In this group, molecules containing Be or Ra are the most polar.

The magnitude and orientation of the permanent electric dipole moments at equilibrium and large interatomic distances correlate with the difference in atomic electronegativities for all the molecules, except for singlet-state CsFr. Electronegativity is a measure of an atom's ability to attract shared electrons to itself. Thus, the permanent electric dipole moments are oriented from more electronegative atoms to less electronegative ones.

The permanent electric dipole moments have been measured for almost all the alkali-metal diatomic molecules consisting of stable isotopes (except KCs) in their ground vibrational level of the ground $X^1\Sigma^+$ electronic state~\cite{DagdigianJCP71,DagdigianJCP72,DeiglmayrPRL08,NiScience08,DeiglmayrPRA10,MolonyPRL14,GuoPRL16}. The RMSE of our calculated values is 0.13$\,$D (4.5$\,$\%). No experimental measurements of permanent electric dipole moments have been reported for other classes of molecules studied in this work.

A part of theoretical studies collected in Tables~\ref{tab:Ref_AM-AM}-\ref{tab:Ref_AEM-AEM} reported calculations of permanent electric dipole moments alongside potential energy curves at different levels of theory, generally in agreement with the present results. Here, we compare our values with the previous most accurate and systematic studies. For the 10 ground-state alkali-metal molecules consisting of stable isotopes, the average absolute differences between the present and previous values are 0.1$\,$D (4.8$\,$\%)~\cite{AymarJCP05}, 0.04$\,$D (2.5$\,$\%)~\cite{FedorovJCP14}, and 0.09$\,$D (4.7$\,$\%)~\cite{MitraPRA20}. For the 16 lightest ground-state alkali-metal--alkaline-earth molecules, the average absolute differences between the present and previous values are 0.15$\,$D (13$\,$\%)~\cite{PototschnigPCCP16} and 0.08$\,$D (6.6$\,$\%)~\cite{MitraPRA22}. The average absolute differences between the present results and values for LiSr, NaSr, KSr, RbSr, and CsSr reported in Ref.~\cite{GueroutPRA10} are 0.08$\,$D (7.9$\,$\%). It is worth mentioning that different methods, basis sets, and pseudopotentials were used in the present work and Refs.~\cite{AymarJCP05,GueroutPRA10,PototschnigPCCP16}. Furthermore, the authors of Refs.~\cite{MitraPRA20,MitraPRA22} included the relativistic effects directly in all-electron calculations in contrast to our scalar relativistic pseudopotentials. Thus the observed overall agreement additionally cross-validates the accuracy of different approaches. To the best of our knowledge, the calculations of permanent electric dipole moments have not been previously reported for alkaline-earth molecules. We estimate the uncertainty of our permanent electric dipole moments to be around~5$\,$\%.

\subsection{Static electric dipole polarizabilities}

The polarizability tensor of  $\Sigma$-state molecules has two independent components: parallel $\alpha^\parallel(R)\equiv\alpha^{zz}(R)$ and perpendicular $\alpha^\perp(R)\equiv\alpha^{xx}(R)=\alpha^{yy}(R)$ ones with $z$ axis chosen along the internuclear axis in a molecule-fixed reference frame. The calculated static electric dipole polarizabilities as functions of the internuclear distance for the alkali-metal diatomic molecules in the $X^1\Sigma^+$ and $a^3\Sigma^+$ electronic states, the alkali-metal--alkaline-earth-metal diatomic molecules in the $X^2\Sigma^+$ electronic state, and the alkaline-earth-metal diatomic molecules in the $X^1\Sigma^+$ state are provided in the Supplemental Material~\cite{supplemental}. Calculations are performed for all the combinations of the alkali-metal (Li, Na, K, Rb, Cs, Fr) and alkaline-earth-metal (Be, Mg, Ca, Sr, Ba, Ra) atoms. The corresponding values for the equilibrium distances, $\alpha^\parallel_e\equiv \alpha^\parallel(R_e)$ and $\alpha^\perp_e\equiv \alpha^\perp(R_e)$, are collected in Tables~\ref{table:AM-AM-S}-\ref{table:AEM-AEM}. The polarizabilities are reported in atomic units of $e^2a_0^2/E_\text{h}$ throughout this paper.

At large internuclear distances, the polarizabilities approach their asymptotic behavior given by the atomic polarizabilities $\alpha_A$ and $\alpha_B$~\cite{JensenJCP02}
\begin{equation}\label{eq:pol_long}
\begin{split}
\alpha^\parallel(R)&\approx \alpha_A+\alpha_B+\frac{4\alpha_A\alpha_B}{R^3}+\frac{4(\alpha_A+\alpha_B)\alpha_A\alpha_B}{R^6}\,,\\
\alpha^\perp(R)&\approx \alpha_A+\alpha_B-\frac{2\alpha_A\alpha_B}{R^3}+\frac{(\alpha_A+\alpha_B)\alpha_A\alpha_B}{R^6}\,.
\end{split}
\end{equation}
Two independent polarizability tensor components can also be transformed to the isotropic $\bar{\alpha}(R)$ and anisotropic $\Delta\alpha(R)$ ones using
\begin{equation}
\begin{split}
\bar{\alpha}(R)&=\frac{2\alpha^\perp(R)+\alpha^\parallel(R)}{3} \,,\\
\Delta\alpha(R)&=\alpha^\parallel(R)-\alpha^\perp(R)\,.
\end{split}
\end{equation}

The polarizability describes the molecular response to an electric field in the second order of perturbation theory. For example, optical dipole trapping is governed by the isotropic polarizability $\bar{\alpha}$, while the laser-induced molecular alignment is controlled by its anisotropic component $\Delta\alpha$~\cite{LemeshkoMP13}. Molecular polarizabilities are also useful in evaluating long-range intermolecular interactions~\cite{JeziorskiCR94}.

The calculated polarizabilities for all the studied molecules are well described by the formulas in Eq.~\eqref{eq:pol_long} at intermediate and large internuclear distances. For the most weakly-bound alkali-metal dimers in the $a^3\Sigma^+$ state, these formulas also apply around their equilibrium distances, while the more strongly-bound alkali-metal dimers in the $X^1\Sigma^+$ state have smaller polarizabilities at equilibrium distances, influenced by stronger chemical bonding of a polarized-covalent nature. In general, the magnitude of the molecular polarizabilities correlates with the atomic polarizabilities, which increase with increasing the atomic number of the alkali-metal and alkaline-earth-metal atoms, with an exception for Fr and Ra due to the contraction of electronic orbitals by the relativistic effects. The parallel components at equilibrium distances are around twice larger than the perpendicular ones for all atomic combinations. 

The polarizabilities have been measured only for a few ground-state alkali-metal molecules~\cite{DagdigianJCP71,DagdigianJCP72,MolofJCP74,KnightPRB85,TarnovskyJCP93,AntoineJCP99,BowlanPRL11,HohmJMS13}. Our results agree with the experimental values within their uncertainties. For example, for LiNa, the calculated isotropic and anisotropic polarizabilities of $\bar{\alpha}_e=237.8\,$ and $\Delta\alpha_e=156.6\,$ can be compared with the experimental ones of $\bar{\alpha}_0=270(34)$ and $\Delta\alpha_0=162(13)$ from Ref.~\cite{GraffJCP72}. Our isotropic values $\bar{\alpha}_e$ of 209.2, 262.7, 483.2, 559.2, 712.6, 357.1, and 589.8 for Li$_2$, Na$_2$, K$_2$, Rb$_2$, Cs$_2$, NaK, and KCs, respectively, agree with the corresponding experimental estimates of 216(20), 256(20), 499(40), 533(40), 702(54), 344(20), and 601(34) from Ref.~\cite{TarnovskyJCP93}. The experimental values for the molecules containing the alkaline-earth-metal atoms are even more scarce. For the Ba$_2$ dimer, the ratio of the polarizabilities of the ground state molecules and separate atoms was measured to be 1.30(13)~\cite{SchaferPRA07}, which agrees with our value of~1.21.  

The present polarizabilities agree well within a few percent with previous systematic computational studies employing different methods, basis sets, and pseudopotentials presented for the alkali-metal molecules in Refs.~\cite{DeiglmayrJCP08,ByrdPRA12,ZuchowskiPRA13,VexiauIRPC17,MitraPRA20} and the alkali-metal--alkaline-earth-metal molecules in Ref.~\cite{GopakumarJCP14,PototschnigPCCP16,MitraPRA22}. The agreement is also satisfactory with other calculations reported in some older works, which used smaller basis sets and less accurate wave functions~\cite{MullerJCP84,MullerJCP86,UrbanJCP95,ReratMP03,DyallJCTC23}.

\section{Summary and conclusions}
\label{sec:summary}

Motivated by the growing experimental interest in producing and using ultracold gases of molecules that contain different alkali-metal or alkaline-earth-metal atoms, we have conducted a comprehensive theoretical study on the ground-state electronic properties of such molecules. We have calculated the interaction energies, permanent electric dipole moments, and static electric dipole polarizabilities as functions of the internuclear distance for all 78 possible homonuclear and homonuclear diatomic combinations of alkali-metal (Li, Na, K, Rb, Cs, Fr) and alkaline-earth-metal (Be, Mg, Ca, Sr, Ba, Ra) atoms. We have employed the hierarchy of coupled cluster methods up to CCSDTQ with large Gaussian basis sets and small-core relativistic energy-consistent pseudopotentials. The inclusion of full triple and quadruple excitations in the coupled cluster method to obtain potential energy curves has allowed for the description of valence electrons at the full configuration interaction level for all molecules. Thus, results for three classes of experimentally relevant molecules have been presented at a consistent level of theory. Corresponding spectroscopic constants have been collected. We have computed the electronic properties of some molecules, such as those containing francium and radium, for the first time. The permanent electric dipole moments of heteronuclear alkaline-earth molecules have also been presented for the first time. We have analyzed the convergence, estimated computational uncertainties, and compared previous experimental and theoretical data with the present values. 

The presented results can serve as a reference and theoretical benchmark for future potentially more accurate electronic-structure computations~\cite{GronowskiPRA20}, including both ground and excited electronic states. The potential energy curves can be used to obtain rovibrational levels to guide and explain the formation and new spectroscopic measurements for experimentally unexplored atomic combinations. Our interaction potentials can also serve as a starting point for fitting curves to future accurate experimental rovibrational energy levels for molecules containing alkaline-earth-metal atoms~\cite{CiameiPCCP18}. 

The calculated permanent electric dipole moments and static electric dipole polarizabilities can be employed to construct long-range intermolecular interaction potentials~\cite{ChristianenJCP19}. They can also be used to describe a molecular response to static electric and off-resonant laser fields~\cite{LemeshkoMP13}, respectively, and, in general, to design and guide new formation, control, and manipulation schemes~\cite{TomzaPRL14,DevolderPRA21}.    

Finally, the determined well depths and related dissociation energies can allow for assessing the energetics of chemical reactions between ground-state molecules and in atom-molecule collisions. The chemical reactivity of the alkali-metal diatomic molecules is well understood~\cite{TscherbulPRA08,ZuchowskiPRA10,ByrdPRA2010,ByrdJCP12,TomzaPRA13a}. On the other hand, the reactivity of the alkali-metal--alkaline-earth-metal (except RbSr~\cite{ManNJP22}) and alkaline-earth-metal molecules have not been theoretically studied, while fast chemical reactions between ground-state Sr$_2$ were recently obsreved~\cite{LeungNJP21}. The presented data can directly provide the energy changes for atom-exchange chemical reactions in molecular gases and atom-molecule mixtures and have been employed in parallel to access the chemical reactivity of alkaline-earth-metal molecules~\cite{LadjimiPRA23}. They can also be used to evaluate the energy changes for trimer formation chemical reactions if binding energies of triatomic molecules are calculated.

Full potential energy curves, permanent electric dipole moments, and static electric dipole polarizabilities as functions of the interatomic distance in a numerical form are collected in the Supplemental Material~\cite{supplemental}.

\begin{acknowledgements}
We gratefully acknowledge the National Science Centre Poland (grant no.~2020/38/E/ST2/00564) for financial support and Poland’s high-performance computing infrastructure PLGrid (HPC Centers: ACK Cyfronet AGH) for providing computer facilities and support (computational grant no.~PLG/2021/015237).
\end{acknowledgements}

\bibliography{XY}

\end{document}